\documentclass[10pt]{article}

\usepackage{graphicx}

\usepackage[all,cmtip]{xy}

\usepackage{setspace}
\usepackage{bbm,youngtab}
\usepackage{amscd,mathrsfs}
\usepackage{amssymb,amsmath,dsfont,amsfonts,amsthm}
\usepackage{euscript,enumerate}
\usepackage{stmaryrd}
\usepackage[cbgreek]{textgreek} 
\usepackage{booktabs}
\usepackage{multirow}
\usepackage{hhline} 
\usepackage{color}

\DeclareMathAlphabet{\mathpzc}{OT1}{pzc}{m}{it}

\allowdisplaybreaks 

\usepackage{authblk}
\usepackage{natbib}
\usepackage{color, soul}
\usepackage{enumerate}
\usepackage{empheq}
\usepackage{rotating}
\usepackage{ftnxtra}
\usepackage{fnpos}
\usepackage[titletoc,toc,title]{appendix}
\usepackage{euscript}
\usepackage{graphicx}
\usepackage{epsfig}
\usepackage{epstopdf}
\DeclareGraphicsExtensions{.pdf,.png,.jpg,.eps}
\usepackage{pstool}
\usepackage{upgreek}
\usepackage{mathrsfs}
\usepackage{tikz-cd}

\usepackage{psfrag}

\numberwithin{equation}{section}

\theoremstyle{plain}	
\newtheorem{thm}{Theorem}[section]

\newtheorem*{prop*}{Proposition}
\theoremstyle{definition}	

\newtheorem{remark}[thm]{Remark}

\usepackage{caption}
\usepackage{subcaption}

\usepackage{hyperref}
\hypersetup{colorlinks=true, linkcolor=blue}
\hypersetup{colorlinks=true,citecolor=blue}

\DeclareMathAlphabet{\mathpzc}{OT1}{pzc}{m}{it}

\usepackage{amsmath, amsthm, amssymb}

\usepackage{cleveref}

\DeclarePairedDelimiter\abs{\lvert}{\rvert}

\makeatletter
\newsavebox{\@brx}
\newcommand{\llangle}[1][]{\savebox{\@brx}{\(\m@th{#1\langle}\)}%
  \mathopen{\copy\@brx\mkern2mu\kern-0.9\wd\@brx\usebox{\@brx}}}
\newcommand{\rrangle}[1][]{\savebox{\@brx}{\(\m@th{#1\rangle}\)}%
  \mathclose{\copy\@brx\mkern2mu\kern-0.9\wd\@brx\usebox{\@brx}}}%
\let\oldabs\abs
\def\abs{\@ifstar{\oldabs}{\oldabs*}}
\makeatother

\usepackage{accents}

\usepackage{mathtools}

    {\end{bmatrix}}%

\usepackage[all,cmtip]{xy}

\usepackage{enumitem}

\usepackage{bbold}

\usepackage{yhmath}

\begin{document}
\bibliographystyle{abbrvnat}

\title{\Large{\textbf{\vspace{-.75in}\\
Geometric Phases of Nonlinear Elastic $N$-Rotors \\via Cartan's Moving Frames
}}}

\author[1]{Francesco Fedele}
\author[1,2]{Arash Yavari\thanks{Corresponding author, e-mail: arash.yavari@ce.gatech.edu}}
\affil[1]{\small \textit{School of Civil and Environmental Engineering, Georgia Institute of Technology, Atlanta, GA 30332, USA}}
\affil[2]{\small \textit{The George W. Woodruff School of Mechanical Engineering, Georgia Institute of Technology, Atlanta, GA 30332, USA}}

\maketitle
\thispagestyle{empty}

\begin{abstract}
\noindent
We study the geometric phases of nonlinear elastic $N$-rotors with continuous rotational symmetry. In the Hamiltonian framework, the geometric structure of the phase space is a principal fiber bundle, i.e., a base, or shape manifold~$\mathcal{B}$, and fibers $\mathcal{F}$ along the symmetry direction attached to it. The symplectic structure of the Hamiltonian dynamics determines the connection and curvature forms of the shape manifold. Using Cartan's structural equations with zero torsion we find an intrinsic (pseudo) Riemannian metric for the shape manifold. \textcolor{black}{ Without lose of generality, we show that one has the freedom to define the rotation sign of the total angular momentum of the elastic rotors as either positive or negative, e.g., counterclockwise or clockwise, respectively, or viceversa. This endows the base manifold~$\mathcal{B}$ with two distinct metrics both compatible with the geometric phase. In particular, the metric is pseudo-Riemannian if $\mathsf{A}<0$, and the shape manifold is a $2$D~Robertson-Walker spacetime with positive curvature.} For $\mathsf{A}>0$, the shape manifold is the hyperbolic plane $\mathbb{H}^2$ with negative curvature. We then generalize our results to free elastic $N$-rotors. We show that the associated shape manifold~$\mathcal{B}$ is reducible to the product manifold of $(N-1)$ hyperbolic planes $\mathbb{H}^2$~($\mathsf{A}>0$), or $2$D~Robertson-Walker spacetimes~($\mathsf{A}<0$) depending on the convection used to define the rotation sign of the total angular momentum. We then consider elastic $N$-rotors subject to time-dependent self-equilibrated moments. \textcolor{black}{The $N$-dimensional shape manifold of the extended autonomous system has a structure similar to that of the $(N-1)$-dimensional shape manifold of free elastic rotors.}
The Riemannian structure of the shape manifold provides an intrinsic measure of the closeness of one shape to another in terms of curvature, or induced geometric phase.  \end{abstract}

\begin{description}
\item[Keywords:]  Geometric phase, Berry's phase, geometric drift, Cartan's moving frames.
\end{description}

\tableofcontents

\section{Introduction} \label{Intro}

A classical example in which geometric phases arise is the parallel transport of a vector tangent to a sphere. The change in the vector direction
is equal to the solid angle of the closed path spanned by the vector and it can be described by Hannay's angles~\citep{Hannay}. The rate at which the angle, or geometric phase, changes in time is the geometric phase velocity. In physics, the rotation of Foucault\textquoteright s pendulum can also be explained by means of geometric phases. \citet{Pancharatnam} discovered their effects in polarized light, and later \citet{Berry1984} rediscovered it for quantum-mechanical systems~(see also~\citep{Berry1990,Simon1983,Aharonov_Anandan,Garrison_GeomPhases}). \citet{Berry1984,Berry1990} showed that a quantum mechanical system that undergoes an adiabatic evolution acquires a phase factor that is purely geometric.

Another example drawn from classical mechanics is the spinning body in a dissipationless medium, which has a rotational symmetry with respect to the axis of rotation. The associated angular, or geometric phase velocity $\Omega$ follows from the conservation of the angular momentum $I\Omega$, where $I$ is the mass moment of inertia. If the body changes shape, $I$ varies over time and so does the angular speed $\Omega$. In the frame rotating at that speed, one only observes the body shape-changing dynamics and the rotational symmetry is reduced. In a fixed frame one cannot distinguish between the body deformation and spinning motion. In general, geometric phases are observed in classical mechanical systems with internal variables that rule their shape deformations, and variables that rule their rigid translation of the system as a whole. A cyclic motion of the shape variables can induce a rigid translation if the total momentum is conserved. 

In classical and quantum mechanics the key geometrical structure is the symplectic form of a Hamiltonian. The Riemannian structure and a metric are traditionally associated to the theories of General Relativity and gravitation. In quantum mechanics, the scalar product on the Hilbert space induces naturally a distance between quantum states, but the interest is not in the local properties of the manifold of states. The physically relevant quantities are transition probability amplitudes between quantum states, which do not depend on their relative distance. However, \citet{ProvostVallee} argued that for macroscopic systems exhibiting collective behaviour, the possibility of going from one state to another is not described by a direct transition amplitude (scalar product in Hilbert space) but rather through a succession of infinitesimal steps on the manifold of collective states. The relevant distance between distinct states is then the distance measured along geodesics on the manifold.

In quantum mechanics, the Riemannian metric is the Fubini-Study metric of complex projective spaces \citep{ProvostVallee,Anandan_metric}. The importance of the associated geodesic curves stems from the fact that Berry's phase between two quantum states can be expressed by integrating the associated connection form along the geodesic between the two states \citep{Bhandari,Wilczek_book}. 
As a matter of fact, the quantum metric provides the infinitesimal distance between two nearby states differing by a Berry phase. Such a distance measures the quantum fluctuations between the two states \citep{ProvostVallee}. 

In fluid mechanics, the motion of a swimmer at low Reynolds numbers can be explained in terms of geometric phases~\citep{Shapere1987,Shapere1}. Swimmers can cyclically change their shape (internal variables) to move forward (translation variables). Since inertia is neglected the swimmer's velocity is uniquely determined by the geometry of the sequence of its body's shapes, which lead to a net translation, i.e., the geometric phase. A fixed observer sees the swimmer drifting as its body shape cyclically changes over time, but it is hard to distinguish between the two motions. On the contrary, an observer moving with the swimmer sees only its body deformations and translation symmetry is reduced in the (symmetry-reduced) moving frame. In wave mechanics, the slowdown of large oceanic wave groups can be explained in terms of geometric phases \citep{FedeleEPL2014,Banner_PRL2014,fedele2020crest}.
Channel flow turbulence governed by the Navier-Stokes equations admits a continuous translation symmetry. Vortical structures, i.e., packets of vorticity, advect downstream at a speed that depends on their intrinsic inertia (dynamical phase) and on the way their $\textit{shape}$ varies over time (geometric phase). \citet{Fedele2015} showed that the geometric phase component of the vortex speed can be interpreted as a $\textit{self-propulsion}$ velocity induced by the shape-changing vortex deformations similar to the motion of a swimmer at low Reynolds numbers~\citep{Shapere1}. 

In the literature, geometric phases have been understood in terms of holonomy of connections on vector bundles \citep{Simon1983}. In this paper we study geometric phases of nonlinear elastic $N$-rotors in the Hamiltonian framework~\citep{Marsden1990} exploiting Cartan's moving frames to characterize the Riemannian structure of the reduced dynamics. We first present a complete analysis of the geometric phases of a coupled elastic double rotor, which conserves total angular momentum. This problem was discussed by \citet{Marsden1990} to introduce the approach of Hamiltonian reduction for  mechanical systems with a continuous Lie symmetry. Such a symmetry implies that the associated phase space has the structure of a principal fiber bundle, i.e., a shape manifold and transversal fibers attached to it. The symplectic form of the Hamiltonian dynamics yields the connection form on the shape manifold, which thus determines the horizontal transport through the fiber bundle. A cyclic flow on the shape manifold induces a drift along the fibers. This includes dynamic and geometric phases. The dynamic phase increases with the time spent by the flow to wander around the phase space and answers the question: ``How long did your trip take?" \citep{Berry1984}. On the contrary, the geometric phase is independent of time and it depends only upon the curvature of the shape manifold, and answers the question: ``Where have you been?"~\citep{Berry1984}. The geometric phase is defined by the connection form. \citet{Marsden1990} defined the associated geometric phases and related them to the curvature form of the shape manifold. Here, we present a new analysis exploiting Cartan's first structural equations with zero torsion and derive the intrinsic Riemannian structure of the shape manifold, which to the best of our knowledge, has not been investigated to this date. The use of Cartan's moving frames in studying the geometric phases of nonlinear elastic $N$-rotors is motivated by the success of the applications of Cartan's machinery in the analysis of distributed defects in nonlinear solids by the second author and co-workers \citep{Yavari2012a,Yavari2012b,Yavari2013,Yavari2014,Yavari2016Dispiration,Golgoon2018}.

This paper is organized as follows. We first review the theory of Cartan's moving frames and associated connection and curvature forms. The theory is then applied to pseudo-Riemannian manifolds. As a special case we derive the Cartan curvature forms of an $N$-dimensional manifold with a diagonal metric. We then introduce the problem of an elastic double rotor in the Hamiltonian setting. The geometric phases of the system are then studied and an intrinsic metric of the shape manifold is derived. We then extend our study to the geometric phases of free nonlinear elastic $N$-rotors and elastic $N$-rotors subject to self-equilibrating external moments. Finally, we discuss the physical relevance of the intrinsic metric for applications, and in particular, to fluid turbulence.

\section{Differential geometry via Cartan's moving frames} \label{Cartan-Movng-Frames}

Given an $N$-manifold $\mathcal{B}$ with a metric $\mathbf{G}$ and an affine connection $\nabla$, $(\mathcal{B},\nabla,\mathbf{G})$ is called a metric-affine manifold \citep{Gordeeva2010}. Here we mainly follow \citet{Hehl2003} and \citet{Sternberg2013}.
Let us consider an orthonormal frame field $\{\mathbf{e}_1(X),\hdots,\mathbf{e}_N(X)\}$ that at every point $X\in\mathcal{B}$ forms a basis for the tangent space $T_X\mathcal{B}$. A moving frame is, in general, a non-coordinate basis for the tangent space. 
The moving frame field $\{\mathbf{e}_{\alpha}\}$ defines the moving co-frame field $\{\vartheta^1,\hdots,\vartheta^N\}$ such that $\vartheta^{\alpha}(\mathbf{e}_{\beta})=\delta^{\alpha}_{\beta}$, where $\delta^{\alpha}_{\beta}$ is the Kronecker delta. 
As the moving frame is assumed to be orthonormal, i.e., $\llangle \mathbf{e}_{\alpha},\mathbf{e}_{\beta}\rrangle_{\mathbf{G}}=\delta_{\alpha\beta}$, where $\llangle .,. \rrangle_{\mathbf{G}}$ is the inner product induced by the metric $\mathbf{G}$, with respect to the moving frame the metric has the representation 
\begin{equation}\label{Cartanmetric}
  \mathbf{G}=\delta_{\alpha\beta}\,\vartheta^{\alpha}\otimes\vartheta^{\beta}\,,   
\end{equation}
where summation over repeated indices is assumed. 

An affine (linear) connection is an operation $\nabla:\mathcal{X}(\mathcal{B})\times\mathcal{X}(\mathcal{B})\rightarrow\mathcal{X}(\mathcal{B})$, where $\mathcal{X}(\mathcal{B})$ is the set of vector fields on $\mathcal{B}$, with certain properties, namely, a) $\nabla_{f_1\mathbf{X}_1+f_2\mathbf{X}_2}\mathbf{Y}=f_1\nabla_{\mathbf{X}_1}\mathbf{Y}+f_2\nabla_{\mathbf{X}_2}\mathbf{Y}$, b) $\nabla_{f_1\mathbf{X}_1+f_2\mathbf{X}_2}\mathbf{Y}=f_1\nabla_{\mathbf{X}_1}\mathbf{Y}+f_2\nabla_{\mathbf{X}_2}\mathbf{Y}$, and c) $\nabla_{\mathbf{X}}(f\mathbf{Y})=f\nabla_{\mathbf{X}}\mathbf{Y}+(\mathbf{X}f)\mathbf{Y}$, where $\mathbf{X}$, $\mathbf{Y}$, $\mathbf{X}_1$, $\mathbf{X}_2$, $\mathbf{Y}_1$, and $\mathbf{Y}_2$ are arbitrary vector fields, $f,f_1,f_2$ are arbitrary functions, and $a_1,a_2$ are arbitrary scalars. The vector $\nabla_{\mathbf{X}}\mathbf{Y}$ is the covariant derivative of $\mathbf{Y}$ along $\mathbf{X}$. 
Given the connection $\nabla$, the connection $1$-forms are defined as
\begin{equation}
    \nabla\mathbf{e}_{\alpha}=\mathbf{e}_{\gamma}\otimes\omega^{\gamma}{}_{\alpha}\,.
\end{equation}
The connection coefficients are defined as $\nabla_{\mathbf{e}_{\beta}}\mathbf{e}_{\alpha}=\left\langle \omega^{\gamma}{}_{\alpha},\mathbf{e}_{\beta} \right\rangle \mathbf{e}_{\gamma}=\omega^{\gamma}{}_{\beta\alpha}\,\mathbf{e}_{\gamma}$.\footnote{$\langle.,.\rangle$ is the natural pairing of $1$-forms and vectors.}
Thus, the connection $1$-forms have the representation $\omega^{\gamma}{}_{\alpha}=\omega^{\gamma}{}_{\beta\alpha}\,\vartheta^{\beta}$. 
It is straightforward to show that $\nabla\vartheta^{\alpha}=-\omega^{\alpha}{}_{\gamma}\,\vartheta^{\gamma}$, and $\nabla_{\mathbf{e}_{\beta}}\vartheta^{\alpha}=-\omega^{\alpha}{}_{\beta\gamma}\,\vartheta^{\gamma}$.

A coordinate chart $\{X^A\}$ for $\mathcal{B}$ defines a coordinate basis $\left\{\partial_A=\frac{\partial}{\partial X^A}\right\}$ for $T_X\mathcal{B}$.
The moving frame field $\{\mathbf{e}_{\alpha}\}$ is related to the coordinate basis by a $GL(N,\mathbb{R})$-rotation: $\mathbf{e}_{\alpha}=\mathsf{F}_{\alpha}{}^A\,\partial_A$. In order to preserve orientation, it is assumed that $\det[\mathsf{F}_{\alpha}{}^A]>0$. 
The relation between the moving and coordinate co-frames is $\vartheta^{\alpha}=\mathsf{F}^{\alpha}{}_A\,dX^A$, where $[\mathsf{F}^{\alpha}{}_A]$ is the inverse of $[\mathsf{F}_{\alpha}{}^A]$.
For the coordinate frame $[\partial_A,\partial_B]=0$, where $[\mathbf{X},\mathbf{Y}]=\mathbf{X}\mathbf{Y}-\mathbf{Y}\mathbf{X}$ is the Lie bracket (commutator) of the vector fields $\mathbf{X}$ and $\mathbf{Y}$. For an arbitrary scalar field $f$, $[\mathbf{X},\mathbf{Y}][f]=\mathbf{X}[f]\mathbf{Y}-\mathbf{Y}[f]\mathbf{X}$.
For the moving frame field one has
\begin{equation}
    [\mathbf{e}_{\alpha},\mathbf{e}_{\beta}]=-c^{\gamma}{}_{\alpha\beta}\,\mathbf{e}_{\gamma}\,,
\end{equation}
where $c^{\gamma}{}_{\alpha\beta}$ are components of the \emph{object of anhonolomy} $c^{\gamma}=d\vartheta^{\gamma}$. 
Noting that 
\begin{equation}
    c^{\gamma}=d\left(\mathsf{F}^{\gamma}{}_B\,dX^B\right)
    =\sum_{\alpha<\beta}c^{\gamma}{}_{\alpha\beta} \,\vartheta^{\alpha}\wedge\vartheta^{\beta}\,,
\end{equation}
one can show that
\begin{equation}
	c^{\gamma}{}_{\alpha\beta}=\mathsf{F}_{\alpha}{}^A\,\mathsf{F}_{\beta}{}^B
	\left(\partial_A\mathsf{F}^{\gamma}{}_B-\partial_B\mathsf{F}^{\gamma}{}_A\right) \,.
\end{equation}
In the local chart $\{X^A\}$, $\nabla_{\partial_A}\partial_B=\Gamma^C{}_{AB}\,\partial_C$, where $\Gamma^C{}_{AB}$ are the Christoffel symbols of the connection. 

\subsection{Non-metricity}

For a metric-affine manifold $(\mathcal{B},\nabla,\mathbf{G})$, non-metricity $\boldsymbol{\mathcal{Q}}:\mathcal{X}(\mathcal{B})\times\mathcal{X}(\mathcal{B})\times\mathcal{X}(\mathcal{B})\rightarrow \mathcal{X}(\mathcal{B})$ is defined as
\begin{equation}
     \boldsymbol{\mathcal{Q}}(\mathbf{X},\mathbf{Y},\mathbf{Z})
     =\llangle \nabla_{\mathbf{X}}\mathbf{Y},\mathbf{Z} \rrangle_{\mathbf{G}}
     +\llangle \mathbf{Y},\nabla_{\mathbf{X}}\mathbf{Z} \rrangle_{\mathbf{G}}
     -\mathbf{X}\big[\llangle\mathbf{Y},\mathbf{Z}\rrangle_{\mathbf{G}}\big]\,.
\end{equation}
In the moving frame $\{\mathbf{e}_{\alpha}\}$, $\mathcal{Q}_{\gamma\alpha\beta}=\boldsymbol{\mathcal{Q}}(\mathbf{e}_{\gamma},\mathbf{e}_{\alpha},\mathbf{e}_{\beta})$. Non-metricity $1$-forms are defined as $\mathcal{Q}_{\alpha\beta}=\mathcal{Q}_{\gamma\alpha\beta}\,\vartheta^{\gamma}$. One can show that $\mathcal{Q}_{\gamma\alpha\beta}
     =\omega^{\xi}{}_{\gamma\alpha}\,G_{\xi\beta}+\omega^{\xi}{}_{\gamma\beta}\,G_{\xi\alpha}
     -\langle dG_{\alpha\beta},\mathbf{e}_{\gamma} \rangle
     =\omega_{\beta\gamma\alpha}+\omega_{\alpha\gamma\beta}
     -\langle dG_{\alpha\beta},\mathbf{e}_{\gamma} \rangle$, where $d$ is the exterior derivative. Hence
\begin{equation}
     \mathcal{Q}_{\alpha\beta}=\omega_{\alpha\beta}+\omega_{\beta\alpha}-dG_{\alpha\beta}\,.
\end{equation}
This is \emph{Cartan's zeroth structural equation}. For an orthonormal frame $G_{\alpha\beta}=\delta_{\alpha\beta}$ and hence
\begin{equation}
     \mathcal{Q}_{\alpha\beta}=\omega_{\alpha\beta}+\omega_{\beta\alpha}.
\end{equation}
The connection $\nabla$ is compatible with the metric $\mathbf{G}$ if non-metricity vanishes, i.e., 
\begin{equation}
    \nabla_{\mathbf{X}}\llangle \mathbf{Y},\mathbf{Z}\rrangle_{\mathbf{G}}
    =\llangle \nabla_{\mathbf{X}}\mathbf{Y},\mathbf{Z} \rrangle_{\mathbf{G}}
    +\llangle \mathbf{Y},\nabla_{\mathbf{X}}\mathbf{Z}\rrangle_{\mathbf{G}}\,.
\end{equation}
This is equivalent to $\nabla\mathbf{G}=\mathbf{0}$, which in a coordinate chart reads $G_{AB|C}=G_{AB,C}-\Gamma^D{}_{CA}G_{DB}-\Gamma^D{}_{CB}G_{AD}=0$. With respect to the moving frame, $\omega_{\alpha\beta}+\omega_{\beta\alpha}=0$, i.e., the connection $1$-forms of a metric-compatible connection are anti-symmetric.

\subsection{Torsion}

Torsion $\boldsymbol{T}:\mathcal{X}(\mathcal{B})\times\mathcal{X}(\mathcal{B})\rightarrow\mathcal{X}(\mathcal{B})$ of the connection $\nabla$ is defined as
\begin{equation}
    \boldsymbol{T}(\mathbf{X},\mathbf{Y})=\nabla_{\mathbf{X}}\mathbf{Y}-\nabla_{\mathbf{Y}}\mathbf{X}
    -[\mathbf{X},\mathbf{Y}]\,.
\end{equation}
In a local chart $\{X^A\}$, torsion has components $T^A{}_{BC}=\Gamma^A{}_{BC}-\Gamma^A{}_{CB}$. 
With respect to the moving frame torsion has the components $T^{\alpha}{}_{\beta\gamma}=\omega^{\alpha}{}_{\beta\gamma}-\omega^{\alpha}{}_{\gamma\beta}+c^{\alpha}{}_{\beta\gamma}$.
The torsion $2$-forms have the following relations with the connection $1$-forms  
\begin{equation}
    \mathcal{T}^{\alpha} = d\vartheta^{\alpha}+\omega^{\alpha}{}_{\beta}\wedge\vartheta^{\beta}\,.
\end{equation}
These are called \emph{Cartan's first structural equations}. 
The connection $\nabla$ is symmetric if it is torsion-free, i.e., $\nabla_{\mathbf{X}}\mathbf{Y}-\nabla_{\mathbf{Y}}\mathbf{X}=[\mathbf{X},\mathbf{Y}]$. With respect to the moving frame, $d\vartheta^{\alpha}+\omega^{\alpha}{}_{\beta}\wedge\vartheta^{\beta}=0$.

\subsection{Curvature}

The curvature $\boldsymbol{\mathcal{R}}:\mathcal{X}(\mathcal{B})\times\mathcal{X}(\mathcal{B})\times\mathcal{X}(\mathcal{B})\rightarrow\mathcal{X}(\mathcal{B})$ of the affine connection $\nabla$ is defined as
\begin{equation}
    \boldsymbol{\mathcal{R}}(\mathbf{X},\mathbf{Y})\mathbf{Z}
    =[\nabla_{\mathbf{X}},\nabla_{\mathbf{Y}}]\mathbf{Z}-\nabla_{[\mathbf{X},\mathbf{Y}]}\mathbf{Z}
    =\nabla_{\mathbf{X}}\nabla_{\mathbf{Y}}\mathbf{Z}
    -\nabla_{\mathbf{Y}}\nabla_{\mathbf{X}}\mathbf{Z}-\nabla_{[\mathbf{X},\mathbf{Y}]}\mathbf{Z}\,.
\end{equation}
In a coordinate chart, $\mathcal{R}^A{}_{BCD}=\Gamma^A{}_{CD,B}-\Gamma^A{}_{BD,C}+\Gamma^A{}_{BM}\,\Gamma^M{}_{CD}-\Gamma^A{}_{CM}\,\Gamma^M{}_{BD}$.
With respect to the moving frame, the curvature tensor has the components $\mathcal{R}^{\alpha}{}_{\beta\lambda\mu}
    =\partial_{\beta}\omega^{\alpha}{}_{\lambda\mu}-\partial_{\lambda}\omega^{\alpha}{}_{\beta\mu}
    +\omega^{\alpha}{}_{\beta\xi}\,\omega^{\xi}{}_{\lambda\mu}
    -\omega^{\alpha}{}_{\lambda\xi}\,\omega^{\xi}{}_{\beta\mu}
    +\omega^{\alpha}{}_{\xi\mu}\,c^{\xi}{}_{\beta\lambda}$.
The curvature $2$-forms are defined as 
\begin{equation} \label{SecondCartan}
    \mathcal{R}^{\alpha}{}_{\beta}=d\omega^{\alpha}{}_{\beta}
    +\omega^{\alpha}{}_{\gamma}\wedge\omega^{\gamma}{}_{\beta}\,.
\end{equation}
These are called \emph{Cartan's second structural equations}. 

Requiring that $\nabla$ be both metric compatible and torsion free determines it uniquely. This is the Levi-Civita connection. With respect to a coordinate chart $\{X^A\}$ it has the connection coefficients (Christoffel symbols) $\Gamma^C{}_{AB}=\frac{1}{2}G^{CD}(G_{BD,A}+G_{AD,B}-G_{AB,D})$. The Levi-Civita connection $1$-forms can be explicitly calculated \citep{ONeill2014}. Using Cartan's first structural equations $d\vartheta^{\alpha}=-\omega^{\alpha}{}_{\beta}\wedge\vartheta^{\beta}$. Thus
\begin{equation}
	d\vartheta^{\alpha}(\mathbf{e}_{\beta},\mathbf{e}_{\gamma})
	=-(\omega^{\alpha}{}_{\beta}\wedge\vartheta^{\beta})(\mathbf{e}_{\beta},\mathbf{e}_{\gamma})
	=-\omega^{\alpha}{}_{\beta}(\mathbf{e}_{\beta})\,\vartheta^{\beta}(\mathbf{e}_{\gamma})
	+\omega^{\alpha}{}_{\beta}(\mathbf{e}_{\gamma})\,\vartheta^{\beta}(\mathbf{e}_{\beta})
	=-\omega^{\alpha}{}_{\beta\gamma}+\omega^{\alpha}{}_{\gamma\beta}\,.
\end{equation}
Similarly,
\begin{equation}
	d\vartheta^{\beta}(\mathbf{e}_{\gamma},\mathbf{e}_{\alpha})
	=-\omega^{\beta}{}_{\gamma\alpha}+\omega^{\beta}{}_{\alpha\gamma}\,,\qquad
	d\vartheta^{\gamma}(\mathbf{e}_{\alpha},\mathbf{e}_{\beta})
	=-\omega^{\gamma}{}_{\beta\alpha}+\omega^{\gamma}{}_{\alpha\beta}\,.
\end{equation}
Thus
\begin{equation}
	d\vartheta^{\alpha}(\mathbf{e}_{\beta},\mathbf{e}_{\gamma})
	+d\vartheta^{\beta}(\mathbf{e}_{\gamma},\mathbf{e}_{\alpha})
	-d\vartheta^{\gamma}(\mathbf{e}_{\alpha},\mathbf{e}_{\beta})
	=2\,\omega^{\alpha}{}_{\gamma\beta}	\,,
\end{equation}
where use was made of the fact that for a metric-compatible connection $\omega^{\alpha}{}_{\gamma\beta}+\omega^{\beta}{}_{\gamma\alpha}=0$. Thus
\begin{equation} \label{Connection-Coefficients}
	\omega^{\alpha}{}_{\gamma\beta}=
	\frac{1}{2}\left[d\vartheta^{\alpha}(\mathbf{e}_{\beta},\mathbf{e}_{\gamma})
	+d\vartheta^{\beta}(\mathbf{e}_{\gamma},\mathbf{e}_{\alpha})
	-d\vartheta^{\gamma}(\mathbf{e}_{\alpha},\mathbf{e}_{\beta})\right]
	\,.
\end{equation}
The components of the Riemann curvature and the Ricci tensor are related to the curvature $2$-forms as
\begin{equation} \label{Riemanntensor}
    \operatorname{Riem}^{\alpha}{}_{\beta\xi\eta}
    =\mathcal{R}^{\alpha}{}_{\beta}(\mathbf{e}_{\xi},\mathbf{e}_{\eta})\,,\qquad 
    \operatorname{Ric}_{\alpha\beta}
    =\mathcal{R}^{\gamma}{}_{\alpha}(\mathbf{\mathbf{e}_\gamma},\mathbf{e}_{\beta})
	\,.
\end{equation}
The Ricci scalar is defined as $\operatorname{R}=\operatorname{Ric}_{\alpha\beta}\delta^{\alpha\beta}$. Note that with respect to the moving frame $g_{\alpha\beta}=\delta_{\alpha\beta}$, and hence $g^{\alpha\beta}=\delta^{\alpha\beta}$.

In the coordinate chart $\{X^A\}$ metric has the components $G_{A B}=\mathsf{F}_{A}{}^{\alpha}\, \mathsf{F}_{B}{}^{\beta}\, \delta_{\alpha\beta}$ and the Riemann and Ricci tensors given in~\eqref{Riemanntensor} have the following components 
\begin{equation}\label{RiemRicciCartanframe}
    \operatorname{Riem}^{A}{}_{B C D}=\mathsf{F}_{\alpha}{}^A \,\mathsf{F}_{B}{}^{\beta}  \,\mathsf{F}_{C}{}^{\xi}\,
    \mathsf{F}_{D}{}^{\eta}\,\operatorname{Riem}^{\alpha}{}_{\beta\xi\eta}\,, \qquad  
    \operatorname{Ric}_{A B}=\mathsf{F}_{A}{}^{\alpha} \,\mathsf{F}_{B}{}^{\beta} \,\operatorname{Ric}_{\alpha\beta}\,,
\end{equation}
where $\mathsf{F}_{A}{}^{\gamma} \,\mathsf{F}_{\gamma}{}^{B}=\delta_A^B$. The Ricci scalar reads $\operatorname{R}=\operatorname{Ric}_{A B} \,G^{A B}$, where $G^{A B}=\mathsf{F}_{\alpha}{}^{A}\, \mathsf{F}_{\beta}{}^{B}\,\delta^{\alpha\beta}$ is the inverse of the metric $G_{A B}$ in the coordinate frame. Since the Ricci scalar is an invariant, its value is the same in any frame. As a matter of fact, $\operatorname{R}=\operatorname{Ric}_{A B} \,G^{A B}
   = \mathsf{F}_{A}{}^{\alpha} \,\mathsf{F}_{B}{}^{\beta} \operatorname{Ric}_{\alpha\beta}\,\mathsf{F}_{\gamma}{}^{A} \,\mathsf{F}_{\rho}{}^{B}\,\delta^{\gamma\rho}
  =(\mathsf{F}_{A}{}^{\alpha} \,
  \mathsf{F}_{\gamma}{}^{A} )(\mathsf{F}_{B}{}^{\beta} \,\mathsf{F}_{\rho}{}^{B})
  \operatorname{Ric}_{\alpha\beta}\delta^{\alpha\rho} 
  =\delta_{\gamma}^{\alpha}\,\delta^{\beta}_{\rho}\,\operatorname{Ric}_{\alpha\beta}\,\delta^{\gamma\rho}
  =\operatorname{Ric}_{\alpha\beta}\,\delta^{\alpha\beta}$.

\subsection{Pseudo-Riemannian manifolds}

For a pseudo-Riemannian manifold, in the Cartan's moving frame the metric $\mathbf{G}$ in \eqref{Cartanmetric} generalizes to~\citep{ONeill2014,Sternberg2013}
\begin{equation}
	\mathbf{G}=\sum_{\alpha=1}^N\epsilon_{\alpha}\,\vartheta^{\alpha}\otimes\vartheta^{\alpha}\,,
\end{equation}
where $\epsilon_{\alpha}=\pm 1$, and $(\epsilon_1,\hdots,\epsilon_N)$ is the signature of the manifold.
The orthonormality of the moving frame field implies that
$\llangle \mathbf{e}_{\alpha},\mathbf{e}_{\beta}\rrangle_{\mathbf{G}}=\delta_{\alpha\beta}\,\epsilon_{\alpha}$ (no summation on $\alpha$).
If the connection is metric compatible, one has 
\begin{equation}
	\omega^{\gamma}{}_{\alpha}\,\delta_{\gamma\beta}\,\epsilon_{\beta}
    +\omega^{\gamma}{}_{\beta}\,\delta_{\gamma\alpha}\,\epsilon_{\alpha}=0\quad (\text{no-summation~on~}\alpha~\text{or~}\beta)
    \,,
\end{equation}
or
\begin{equation}
    \omega_{\alpha\beta}+\omega_{\beta\alpha}=0 \,.
\end{equation}
Thus
\begin{equation}
	\epsilon_{\alpha}\,\omega^{\alpha}{}_{\beta}
    +\epsilon_{\beta}\,\omega^{\beta}{}_{\alpha}=0\quad (\text{no-summation~on~}\alpha~\text{or~}\beta) \,,
\end{equation}
which is equivalent to
\begin{equation}\label{omegaalphabeta}
	\omega^{\alpha}{}_{\beta}=-
    \epsilon_{\alpha}\,\epsilon_{\beta}\,\omega^{\beta}{}_{\alpha}\quad (\text{no-summation~on~}\alpha~\text{or~}\beta) \,.
\end{equation}
The first and the second structural equations remain unchanged. The expressions for the Riemann and Ricci curvatures remain unaltered as well.
The Ricci scalar has the following expression 
\begin{equation}
	 \operatorname{R}=\operatorname{Ric}_{\alpha\beta}G^{\alpha\beta}=\sum_{\alpha=1}^N\operatorname{Ric}_{\alpha\alpha}\,\epsilon_{\alpha}\,.
\end{equation}

\subsection{Riemannian product spaces}
\label{Sec:ProductSpace}

Let $(\mathcal{B}_1,\mathbf{G}_1)$, ..., and $(\mathcal{B}_{N},\mathbf{G}_N)$ be Riemannian manifolds and $\mathcal{B}_1\times \cdots\times \mathcal{B}_N$ be their product manifold. At any point $(X_1,\cdots,X_N)\in\mathcal{B}_1\times\cdots\times \mathcal{B}_N$, one has the direct sum $T_{(X_1,\hdots,X_N)}(\mathcal{B}_1\times\cdots\times \mathcal{B}_N)\cong T_{X_1}\mathcal{B}_1\oplus\cdots\oplus T_{X_N}\mathcal{B}_N$, where $\cong$ means ``isomorphic to". The product metric $\mathbf{G}_1\times\cdots\times \mathbf{G}_N$ on $\mathcal{B}_1\times \cdots\times\mathcal{B}_N$ is defined as
\begin{equation}
	\mathbf{G}_1\times\cdots\times \mathbf{G}_N\big|_{(X_1,\hdots,X_N)}
	=\mathbf{G}_1\big|_{X_1}+\cdots+\mathbf{G}_N\big|_{X_N}
	\,,\quad\forall X_1\in\mathcal{B}_1, \cdots,X_N\in\mathcal{B}_N\,.
\end{equation}
The Riemannian manifold $(\mathcal{B}_1\times\cdots\times \mathcal{B}_N,\mathbf{G}_1\times\cdots\times \mathbf{G}_N)$ is called a Riemannian product space \citep{Joyce2007}. If a Riemannian manifold is isometric to a Riemannian product space, it is called reducible (decomposable). Otherwise, it is irreducible (indecomposable). 
It should be noted that for a Riemannian product space, 
\begin{equation}
	\nabla^{\mathbf{G}_1\times\cdots\times \mathbf{G}_N}_{(\mathbf{U}_1,\cdots,\mathbf{U}_N)}(\mathbf{W}_1,
	\cdots,\mathbf{W}_N)=\left(\nabla^{\mathbf{G}_1}_{\mathbf{U}_1}\mathbf{W}_1,\cdots,
	\nabla^{\mathbf{G}_N}_{\mathbf{U}_N}\mathbf{W}_N\right)\,.
\end{equation}
In particular, the Ricci curvature of the Riemannian product space is written as
\begin{equation}
	\operatorname{Ric}\left((\mathbf{U}_1,\cdots,\mathbf{U}_N),(\mathbf{W}_1,\cdots,\mathbf{W}_N)\right)
	=\operatorname{Ric}_1(\mathbf{U}_1,\mathbf{W}_1)+\cdots+\operatorname{Ric}_N(\mathbf{U}_N,\mathbf{W}_N)\,.
\end{equation}

\subsection{Cartan's curvature $2$-forms of an $N$-dimensional pseudo-Riemannian manifold with a diagonal metric}

Consider an $N$-dimensional pseudo-Riemannian manifold $\mathcal{B}$ that in a coordinate chart $\{X^A\}$ has a diagonal metric
\begin{equation}
	\mathbf{G}=\sum_{A=1}^N \epsilon_{A}\,\mathsf{G}_{A}\,dX^A\otimes dX^A
	=\sum_{A=1}^N \epsilon_{A}\,\sqrt{\mathsf{G}_{A}}\,dX^A\otimes \sqrt{\mathsf{G}_{A}}\,dX^A\,,
\end{equation}
where $\mathsf{G}_{A}\geq 0$ with at least one being positive.
Let us define the co-frame field 
\begin{equation}
	E^*=\left\{\vartheta^A=\sqrt{\mathsf{G}_{A}}\,dX^A\right\} \quad (\text{no~summation~on} ~A) \,,
\end{equation}
and its dual moving frame field 
\begin{equation}
	E=\left\{\mathbf{e}_A= \frac{1}{\sqrt{\mathsf{G}_{A}}}\,\partial_A \right\} \quad (\text{no~summation~on} ~A) \,, 
\end{equation}
which by construction $\vartheta^A(\mathbf{e}_B)=\delta_B^A$. Then, the metric in the moving frame $E$ is simply written as
\begin{equation}
	\mathbf{G}=\sum_{A=1}^N \epsilon_{A}\,\vartheta^A\otimes \vartheta^A\,.
\end{equation}
Note that
\begin{equation}\label{dtheta}
	d\vartheta^A=\sum_{B=1}^N \frac{\partial_B \mathsf{G}_{A}}{2\mathsf{G}_{A}\sqrt{\mathsf{G}_{B}}} 
	\,\vartheta^B\wedge \vartheta^A \quad (\text{no~summation~on} ~A) \,.
\end{equation}
We next calculate the Levi-Civita connection $1$-forms for which $\mathcal{T}^{A}=0$. Note that $\omega_{BA}=-\omega_{AB}$ and there are $N(N-1)/2$ connection $1$-forms to be determined. Cartan's first structural equations read $d\vartheta^A+\omega^A{}_B\,\wedge \vartheta^B=0$. Note that one can use \eqref{Connection-Coefficients}. However, there is an easier approach for calculating the connection $1$-forms.
Recalling that $\omega^A{}_B=\omega^A{}_{CB}\,\vartheta^C$, we have
\begin{equation}
	\sum_{B=1}^N \frac{\mathsf{G}_{A,B}}{2\mathsf{G}_{A}\sqrt{\mathsf{G}_{B}}} 
	\,\vartheta^B\wedge \vartheta^A+ \omega^A{}_{CB}\,\vartheta^C\wedge\vartheta^B=0
	\quad (\text{no~summation~on} ~A) \,,
\end{equation}
where $\mathsf{G}_{A,B}=\partial_B \mathsf{G}_{A}$. Thus
\begin{equation}
	\sum_{B=1}^N 
	\left(\omega^A{}_{CB}\,\vartheta^C
	-\frac{\mathsf{G}_{A,B}}{2\mathsf{G}_{A}\sqrt{\mathsf{G}_{B}}} \,\vartheta^A\right)
	\wedge\vartheta^B=0
	\quad (\text{no~summation~on} ~A) \,.
\end{equation}
Cartan's lemma implies that \citep{Sternberg1999}
\begin{equation}
	\omega^A{}_{CB}\,\vartheta^C
	-\frac{\mathsf{G}_{A,B}}{2\mathsf{G}_{A}\sqrt{\mathsf{G}_{B}}} \,\vartheta^A
	=\xi^A{}_{BC}\,\vartheta^C
	\quad (\text{no~summation~on} ~A~\text{or}~B) \,,
\end{equation}
where $\xi^A{}_{BC}(X)=\xi^A{}_{CB}(X)$ are $\frac{N^2(N+1)}{2}$ arbitrary functions. Thus
\begin{equation}
	\omega^A{}_{B}=\frac{\mathsf{G}_{A,B}}{2\mathsf{G}_{A}\sqrt{\mathsf{G}_{B}}} \,\vartheta^A
	+\xi^A{}_{BC}\,\vartheta^C
	\quad (\text{no~summation~on} ~A~\text{or}~B) \,.
\end{equation}
Hence
\begin{equation}
	\omega_{AB}=\epsilon_{A}\,\frac{\mathsf{G}_{A,B}}{2\mathsf{G}_{A}\sqrt{\mathsf{G}_{B}}} \,\vartheta^A
	+\epsilon_{A}\,\xi^A{}_{BC}\,\vartheta^C
	\quad (\text{no~summation~on} ~A~\text{or}~B) \,.
\end{equation}
Knowing that $\omega_{AB}+\omega_{BA}=0$, one can guess that
\begin{equation}
	\omega_{AB}=\epsilon_{A}\,\frac{\mathsf{G}_{A,B}}{2\mathsf{G}_{A}\sqrt{\mathsf{G}_{B}}} \,\vartheta^A
	-\epsilon_{B}\,\frac{\mathsf{G}_{B,A}}{2\mathsf{G}_{B}\sqrt{\mathsf{G}_{A}}} \,\vartheta^B
	\quad (\text{no~summation~on} ~A~\text{or}~B) \,.
\end{equation}
Thus
\begin{equation} \label{Levi-Civita-connection-forms}
	\omega^A{}_{B}=\mathsf{L}_{A B} \,\vartheta^A
	-\epsilon_{A}\,\epsilon_{B}\,\mathsf{L}_{B A} \,\vartheta^B
	\quad (\text{no~summation~on} ~A~\text{or}~B) \,,
\end{equation}
where
\begin{equation}
\mathsf{L}_{A B}=\frac{\mathsf{G}_{A,B}}{2\mathsf{G}_{A}\sqrt{\mathsf{G}_{B}}}\,.
\end{equation}
It is straightforward to check that the $1$-forms given in \eqref{Levi-Civita-connection-forms} satisfy Cartan's first structural equations, and hence, are the unique Levi-Civita connection $1$-forms.
Note that $\omega^{A}{}_{B}=-\epsilon_{A}\,\epsilon_{B}\,\omega^{B}{}_{A}$, and 
\begin{equation}
\begin{aligned}
    d\omega^{A}{}_{B} &=\sum_C(\mathsf{L}_{A B,C}+\mathsf{L}_{A B}\mathsf{L}_{A C} )\,\vartheta^C\wedge \vartheta^A \\
	& \quad -\sum_C\epsilon_{A}\,\epsilon_{B}\,(\mathsf{L}_{B A,C}+\mathsf{L}_{B A} \mathsf{L}_{B C}) \,\vartheta^C\wedge \vartheta^B
	\quad (\text{no~summation~on} ~A~\text{or}~B) \,,
\end{aligned}
\end{equation}
where from \eqref{dtheta} the relation $d\vartheta^A=\mathsf{L}_{A C} \,\vartheta^C\wedge \vartheta^A$ (no summation on $A$) has been used. 

Cartan's second structural equations read
\begin{equation} \label{SecondCartan}
	\mathcal{R}^A{}_B=d\omega^A{}_B+\omega^A{}_C\wedge\omega^C{}_B\,,  
\end{equation}
where curvature $2$-forms are (pseudo) anti-symmetric, i.e., $\mathcal{R}^A{}_B+\epsilon_A\,\epsilon_B \,\mathcal{R}^B{}_A=0$. More explicitly, 
\begin{equation} \label{SecondCartan2}
\begin{aligned}
	\mathcal{R}^A{}_B &= \sum_C(\mathsf{L}_{A B,C}+\mathsf{L}_{A B}\mathsf{L}_{A C}-\mathsf{L}_{A C}\mathsf{L}_{C B} )\,\vartheta^C\wedge \vartheta^A \\
 &  -\sum_C \epsilon_{A}\,\epsilon_{B}\,(\mathsf{L}_{B A,C}+\mathsf{L}_{B A}\mathsf{L}_{B C}-\mathsf{L}_{B C}\mathsf{L}_{C A}) \,\vartheta^C\wedge \vartheta^B	\\
 &  -\sum_C \epsilon_C\,\epsilon_B\, \mathsf{L}_{A C}\,\mathsf{L}_{B C}\,\vartheta^A\wedge\vartheta^B \qquad \qquad
 (\text{no~summation~on} ~A~\text{or}~B) \,.
 \end{aligned}
\end{equation}

\subsection{Curvature $2$-forms of a $2$-dimensional pseudo-Riemannian manifold with a diagonal metric}

Consider a two-dimensional pseudo-Riemannian manifold and a coordinate chart $U=\{X^1,X^2\}$. Assume a  diagonal metric in the coordinate frame
\begin{equation}\label{metricFF}
	\mathbf{G}=\epsilon_1\,\mathsf{G}_1\,d X^1\otimes d X^1+\epsilon_2\,\mathsf{G}_2\,d X^2\otimes d X^2\,,
\end{equation}
where $\mathsf{G}_1, \mathsf{G}_2\geq 0$, and $\mathsf{G}_1+\mathsf{G}_2 > 0$.
From \eqref{Levi-Civita-connection-forms}, there is only one connection $1$-form $\omega^1{}_2$ given as
\begin{equation} \label{connectionforms}
	\omega^1{}_2=\frac{\mathsf{G}_{1,2}}{2\sqrt{\mathsf{G}_{1}\mathsf{G}_{2}}}\,dX^1
	-\epsilon_1\epsilon_2\,\frac{\mathsf{G}_{2,1}}{2\sqrt{\mathsf{G}_{1}\mathsf{G}_{2}}}\,dX^{2}\,,
\end{equation}
where $\mathsf{G}_{1,2}=\partial_{X_2}\mathsf{G}_{1}$. Alternatively,
\begin{equation} \label{connectionforms2}
	\omega^1{}_2=\frac{\mathsf{G}_{1,2}}{2\mathsf{G}_{1}\sqrt{\mathsf{G}_{2}}}\,\vartheta^1
	-\epsilon_1\epsilon_2\,\frac{\mathsf{G}_{2,1}}{2\mathsf{G}_{2}\sqrt{\mathsf{G}_{1}}}\,\vartheta^{2}\,.
\end{equation}
From Cartan's second structural equations ~\eqref{SecondCartan}, there is only one curvature $2$-form $\mathcal{R}^1{}_2$, which reads
\begin{equation}\label{R12curvatureform}
	\mathcal{R}^1{}_2=d\omega^1{}_2
	=-\frac{1}{2}\left[\epsilon_1\epsilon_2\left(\frac{\mathsf{G}_{2,1}}{\sqrt{\mathsf{G}_{1}\mathsf{G}_{2}}}\right)_{,1}
	+\left(\frac{\mathsf{G}_{1,2}}{\sqrt{\mathsf{G}_{1}\mathsf{G}_{2}}}\right)_{,2}\right]dX^{1}\wedge dX^{2}\,.
\end{equation}
Alternatively, 
\begin{equation}
	\mathcal{R}^1{}_2=K\, \vartheta^{1}\wedge \vartheta^{2}\,,
\end{equation}
where the Gaussian curvature $K=\mathcal{R}^1{}_2(\mathbf{e}_1,\mathbf{e}_2)$ is written as
\begin{equation}\label{Gaussiancurvature}
 	K=-\frac{1}{2\sqrt{\mathsf{G}_{1}\mathsf{G}_{2}}}
	\left[\epsilon_1\epsilon_2 \left(\frac{\mathsf{G}_{2,1}}{\sqrt{\mathsf{G}_{1}\mathsf{G}_{2}}}\right)_{,1}
	+\left(\frac{\mathsf{G}_{1,2}}{\sqrt{\mathsf{G}_{1}\mathsf{G}_{2}}}\right)_{,2}\right]\,.
\end{equation}
From \eqref{omegaalphabeta} it follows that $d\omega^1{}_2=-\epsilon_1\epsilon_2\, d\omega^2{}_1$ and hence
\begin{equation}
  \mathcal{R}^2{}_1=-\epsilon_1\epsilon_2\,\mathcal{R}^1{}_2\,.
\end{equation}
The Ricci tensor is calculated using~\eqref{Riemanntensor}$_2$ as
\begin{equation}\label{Riccitensor2d}
	\operatorname{Ric}_{\alpha\beta}
	=\mathcal{R}^{\gamma}{}_{\alpha}(\mathbf{e}_\gamma,\mathbf{e}_{\beta})
	=\mathcal{R}^{1}{}_{\alpha}(\mathbf{e}_1,\mathbf{e}_{\beta})
	+\mathcal{R}^{2}{}_{\alpha}(\mathbf{e}_2,\mathbf{e}_{\beta})
	\,.
\end{equation}
In particular, 
\begin{equation}
 \operatorname{Ric}_{11} =\mathcal{R}^{2}{}_{1}(\mathbf{e}_2,\mathbf{e}_{1})=\epsilon_1\epsilon_2\, \mathcal{R}^{1}{}_{2}(\mathbf{e}_1,\mathbf{e}_2)=\epsilon_1\epsilon_2\,K\,,\qquad  \operatorname{Ric}_{22}=\mathcal{R}^{1}{}_{2}(\mathbf{e}_1,\mathbf{e}_{2})=K\,,
\end{equation}
and $\operatorname{Ric}_{12}=\operatorname{Ric}_{21}=0$. The Ricci scalar is calculated as
\begin{equation}\label{Ricciscalar2D}
   \operatorname{R}=\epsilon_1\operatorname{Ric}_{11}+\epsilon_2\operatorname{Ric}_{22}=\epsilon_1^2 \epsilon_2 K+\epsilon_2 K = 2\epsilon_2 K=-\frac{1}{\sqrt{\mathsf{G}_{1}\mathsf{G}_{2}}}
	\left[\epsilon_1 \left(\frac{\mathsf{G}_{2,1}}{\sqrt{\mathsf{G}_{1}\mathsf{G}_{2}}}\right)_{,1}
	+\epsilon_2\left(\frac{\mathsf{G}_{1,2}}{\sqrt{\mathsf{G}_{1}\mathsf{G}_{2}}}\right)_{,2}\right]\,. 
\end{equation}
 \textcolor{black}{
 Note that for a pseudo-Riemannian metric in the moving frame, $G_{\alpha\beta}=\epsilon_{\alpha}\delta_{\alpha\beta}$ (no summation on $\alpha$). It follows that $\frac{1}{2}\operatorname{R} G_{12}=\frac{1}{2}\operatorname{R} G_{21}=0$. Also, notice that $\frac{1}{2}\operatorname{R} G_{11}=\epsilon_2\,K\,\epsilon_1=\epsilon_1\,\epsilon_2\,K$, and $\frac{1}{2}\operatorname{R} G_{22}=\epsilon_2\,K\,\epsilon_2=K$. Thus, we have shown that
\begin{equation}
\operatorname{Ric}_{\alpha\beta}-\frac{1}{2}\operatorname{R} G_{\alpha\beta}=0\,,
\end{equation}
i.e., $\mathbf{G}$ in \eqref{metricFF} is the metric of an Einstein manifold~\citep{Besse:Besse1987}.
}

\section{Dynamics of (free) nonlinear elastic double rotor}

In this section we study the geometric phase of a coupled elastic double rotor, which conserves total angular momentum. A similar problem was discussed by~\citet{Marsden1990} to introduce the Hamiltonian reduction technique for mechanical systems with symmetries. The continuous symmetry implies that the associated phase space has the structure of a principal fiber bundle, i.e., a shape manifold and transversal fibers attached to it. \citet{Marsden1990} defined the associated geometric phases and related them to the curvature form of the shape manifold. Hereafter, we present a new analysis exploiting Cartan's structural equations with zero torsion and find the Riemannian structure of the shape manifold, which was not investigated in ~\citep{Marsden1990}. 

\begin{figure}[h!]
\vskip 0.2in
\centering
\includegraphics[width=0.5\textwidth]{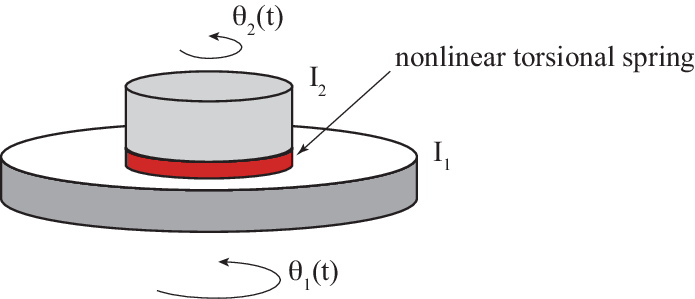}
\vskip 0.1in
\caption{An elastic double rotor with a nonlinear spring. 
}
\label{Fig1}
\end{figure}

Consider the elastic double rotor depicted in Fig.~\ref{Fig1}. The associated Lagrangian is written as
\begin{equation} \label{Lagr}
    \mathcal{L}=\frac{1}{2}I_{1}\,\dot{\theta}_{1}^{2}
    +\frac{1}{2}I_{2}\,\dot{\theta}_{2}^{2}-\Pi(\theta_1,\theta_2)\,,
\end{equation}
where the Lagrangian coordinates $\theta_j$ are the angular positions of the two rotors with mass moments of inertia $I_1$ and $I_2$ as indicated in Fig.~\ref{Fig1}. The potential $\Pi(\theta_1,\theta_2)$ describes conservative moments $M_j=-\partial_{\theta_j}\Pi$, which are in equilibrium, that is
\begin{equation} \label{balance1}
    M_1+M_2=-\frac{\partial \Pi}{\partial \theta_1}-\frac{\partial \Pi}{\partial \theta_2}=0\,. 
\end{equation}
Thus, the potential must be a function of the Lagrangian coordinate difference, i.e., $\Pi=\Pi(\theta_2-\theta_1)$, which is the potential of a nonlinear spring, see Fig.~\ref{Fig1}. Extremizing the action $\int \mathcal{L} dt $ yields the following dynamical equations
\begin{equation} \label{Lagrangeeqs}
 	\frac{d}{dt}\left(\frac{\partial\mathcal{L}}{\partial\dot{\theta}_j}\right)
	-\frac{\partial\mathcal{L}}{\partial\theta_{j}}=I_j\ddot{\theta}_j+\frac{\partial \Pi}{\partial\theta_j}=0\,,
	\quad j=1,2\,.   
\end{equation}
From~\eqref{balance1} potential moments are in equilibrium and summing up equations~\eqref{Lagrangeeqs} yields 
\begin{equation} 
	I_{1}\ddot{\theta}_{1}+I_{2}\ddot{\theta}_{2}=M_1+M_2=0\,.
\end{equation}
Thus, the total angular momentum 
\begin{equation}\label{angulmomem}
 \mathsf{A}=I_{1}\dot{\theta}_{1}+I_{2}\dot{\theta}_{2}\,,
\end{equation}
is conserved. In the following, we assume that $\mathsf{A}\neq 0$. Such an invariant endows the system with a continuous Lie symmetry: if the pair $\mathbf{Z}=(\theta_1(t),\theta_2(t))$ is a solution of the Lagrangian equations, so is 
\begin{equation}\label{Liesymmetry}
    G_{\beta}(\mathbf{Z})=(\theta_1(t)+\beta,\theta_2(t)+\beta)\,,
\end{equation}
 for any angle $\beta\in\mathbb{R}$. In the following, we will use this symmetry in the Hamiltonian setting to reveal the geometric structure of the phase space as that of a principal fiber bundle. Then, the associated Riemannian structure follows from Cartan's structural equations as described in \S\ref{Cartan-Movng-Frames}. 

\subsection{The Hamiltonian structure}

The conjugate momenta follow from the Lagrangian~\eqref{Lagr} as
\begin{equation}
	p_{j}=\frac{\partial\mathcal{L}}{\partial\dot{\theta}_{j}}=I_{j}\dot{\theta}_{j}\,,\qquad j=1,2\,,
\end{equation}
and $\theta_j=p_j/I_j$. Then, the Legendre transform of $\mathcal{L}$ gives the Hamiltonian
\begin{equation} \label{Hamil}
	\mathcal{H}=p_1 \,\dot{\theta}_1+p_2\, \dot{\theta}_2 -\mathcal{L}
	=\frac{1}{2}\frac{p_{1}^{2}}{I_{1}}+\frac{1}{2}\frac{p_{2}^{2}}{I_{2}}
	+\Pi\left(\theta_2-\theta_1\right)\,.
\end{equation}
The configuration space is a $2$-torus  $Q=\mathbb{T}^2$, which has the local chart $\{\theta_1,\theta_2\}$, and the phase space is $T^{*}Q$ with local coordinates $\{\theta_1,\theta_2, p_1,p_2\}$, where $T^{*}Q$ is the cotangent bundle of $Q$. Let us define the vector 
\begin{equation} 
	\mathbf{X}=
 \begin{bmatrix}
 \theta_{1} \\ 
 \theta_{2} \\ 
 p_{1} \\ 
 p_{2}
 \end{bmatrix}\,.
\end{equation}
The dynamics is governed by
\begin{equation}\label{hamildyn}
	\dot{\mathbf{X}}=\mathbf{J}\nabla_{\mathbf{X}}\mathcal{H}\,,
\end{equation}
where 
\begin{equation} 
	\nabla_{\mathbf{X}}=
 \begin{bmatrix}
 \partial_{\theta_1} \\ \partial_{\theta_2} \\
 \partial_{p_1} \\
 \partial_{p_2}
 \end{bmatrix}\,,
\end{equation}
and $\mathbf{J}$ is the following $4\times4$ symplectic matrix
\begin{equation}
    \mathbf{J}=\left[\begin{array}{cc}
	\mathbf{O}_{2} & \mathbf{I}_{2}\\
	-\mathbf{I}_{2} & \mathbf{O}_{2}
	\end{array}\right]=\left[\begin{array}{cccc}
	0 & 0 & 1 & 0\\
	0 & 0 & 0 & 1\\
	-1 & 0 & 0 & 0\\
	0 & -1 & 0 & 0
\end{array}\right]\,.
\end{equation}
$\mathbf{I}_2=[\delta_{ij}]$ is the $2\times2$ identity matrix, $\mathbf{O}_{2}$ is the $2\times2$ null matrix, and $\delta_{ij}$ is the Kronecker tensor. From \eqref{hamildyn}, 
\begin{equation}
	\dot{\theta}_j=\frac{\partial\mathcal{H}}{\partial p_j}\,,\qquad
	\dot{p}_j=-\frac{\partial\mathcal{H}}{\partial\theta_j}\,,\qquad j=1,2\,,
\end{equation}
or 
\begin{equation}\label{Hameq}
	\dot{\theta}_{j}=\frac{p_{j}}{I_{j}}\,,\qquad\dot{p}_{j}=-\frac{\partial \Pi}{\partial\theta_{j}}\,,\qquad j=1,2\,.
\end{equation}

The Hamiltonian $\mathcal{H}$ and the total angular momentum $\mathsf{A}=p_1+p_2$ given in~\eqref{angulmomem} are invariants of motion. The Hamiltonian system inherits the continuous Lie-group symmetry in~\eqref{Liesymmetry}, that is 
\begin{equation}\label{Liesymmetry2}
    G_{\beta}(\mathbf{X})=(\theta_1+\beta,\theta_2+\beta,p_1,p_2),  
\end{equation}
for any angle $\beta\in\mathbb{R}$. The associated $1$-form is
\begin{equation} \label{1form}
	\alpha=p_1 d\theta_1 + p_2 d\theta_2\,,
\end{equation}
and the symplectic $2$-form is defined as
\begin{equation} \label{2form}
	d\alpha=dp_1\wedge d\theta_1 + dp_2\wedge d\theta_2\,.
\end{equation}

\vskip 0.2in
\begin{figure}[h!]
\centering
\includegraphics[width=0.5\textwidth]{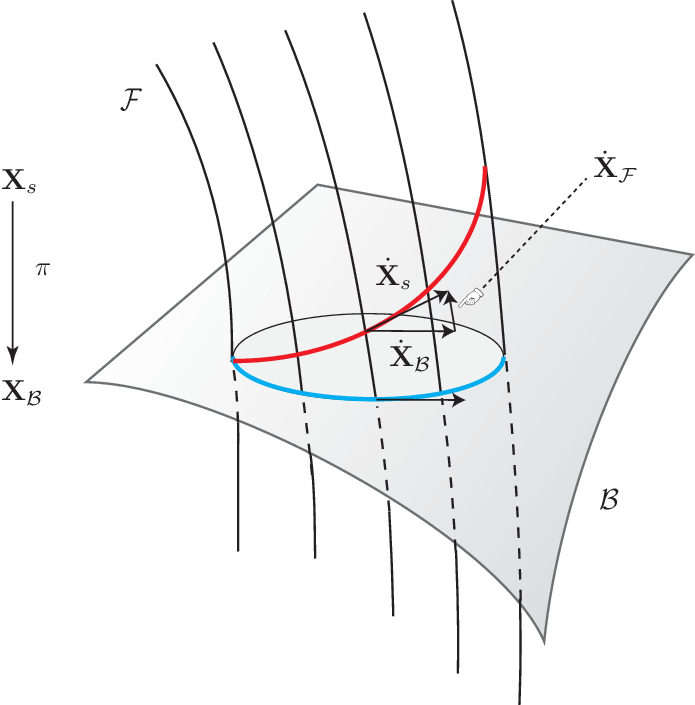}
\vskip 0.0in
\caption{Fiber bundle structure of the state space $\mathcal{P}=T^{*}Q/\mathsf{A}$. $\mathcal{B}$ is the base (shape) manifold and $\mathcal{F}$ is a generic fiber.}
\label{Fig2}
\end{figure}

\subsection{Hamiltonian reduction and geometric phases}

To reveal the geometric nature of the dynamics, we consider another configuration space $Q_s$ with Lagrangian coordinates $\{\theta_1,\psi_2\}$, where the shape parameter~$\psi_2=\theta_2-\theta_1$ represents the relative angular displacement of the two rotors. Since the total angular momentum $\mathsf{A}=p_1+p_2$ must be conserved, $p_1=\mathsf{A}-p_2$ and the motion must occur on the subspace $T^{*}Q_s/\mathsf{A}$ with the coordinate chart $\{\theta_1,\psi_2,p_2\}$. 

The $1$-form in \eqref{1form} reduces to
\begin{equation}
	\alpha=p_1 \,d\theta_1 + p_2 \,d\theta_2=(\mathsf{A}-p_2) \,d\theta_1 + p_2 \,d\theta_2\,,
\end{equation}
and since $\psi_2=\theta_2-\theta_1$, one has
\begin{equation} \label{1formred}
	\alpha=\mathsf{A} \,d\theta_1 + p_2 \,d\psi_2\,.
\end{equation}
The associated symplectic $2$-form reads
\begin{equation}
	d\alpha= dp_2\wedge d\psi_2\,. 
\end{equation}

The reduced state space $\mathcal{P}=T^*Q_s/\mathsf{A}$ has the geometric structure of a principal fiber bundle characterized by the quadruplet $(\mathcal{P},\mathcal{B},G_{\alpha},\pi)$. The flow $\dot{\mathbf{X}}_s\in TQ_s$ in the state space~$\mathcal{P}$ can be decomposed as the sum $\dot{\mathbf{X}}_s=\dot{\mathbf{X}}_{\mathcal{B}}+\dot{\mathbf{X}}_{\mathcal{F}}$
of a Hamiltonian flow 
\begin{equation}
	\mathbf{\dot{X}}_{\mathcal{B}}=\begin{bmatrix}
    \dot{p}_2\\ \dot{\psi_2}
\end{bmatrix}\in T\mathcal{B}\,, 
\end{equation}
on a two-dimensional shape, or base manifold~$\mathcal{B}$ with the coordinate chart $\{p_2,\psi_2\}$, and a drift flow $\mathbf{\dot{X}}_{\mathcal{F}}=\big[\dot{\theta}_1\big]\in T\mathcal{F}$ along the one-dimensional fibers~$\mathcal{F}$ with the coordinate chart $\{\theta_1\}$ as depicted in Fig.~\ref{Fig2}. 
The map $\pi:\mathcal{P}\overset{}{\rightarrow}\mathcal{S}$
projects an element $\mathbf{X}_s$ of the state space $\mathcal{P}$ and all
the elements of the fiber, or  group orbit $G_{\alpha}(\mathbf{X}_s)$, into the same point
$\pi(\mathbf{X}_s)$ of the base manifold $\mathcal{B}$, viz. $\pi(\mathbf{X}_s)=\pi(G_{\beta}(\mathbf{X}_s))$, with $\beta\in\mathbb{R}$. In particular, 
\begin{equation}
	\pi(\mathbf{X}_s)=\mathbf{X}_{\mathcal{B}}=\begin{bmatrix}
    p_2 \\ \psi_2
\end{bmatrix}\,. 
\end{equation}
The reduced Hamiltonian flow $\dot{\mathbf{X}}_{\mathcal{B}}$ on $\mathcal{B}$ is governed by the following equation
\begin{equation} \label{dynred}
	\dot{\mathbf{X}}_{\mathcal{B}}=\mathbf{J}_R\nabla_{\mathbf{X}_\mathcal{B}}\mathcal{H}_R\,,
\end{equation}
where the reduced Hamiltonian is written as
\begin{equation} \label{Hamilreduced}
	\mathcal{H}_R=\frac{1}{2}\frac{\left(p_2-\mathsf{A}\right)^2}{I_1}+\frac{1}{2}\frac{p_2^2}{I_2}
	+\Pi\left(\psi_2\right)\,.
\end{equation}
$\mathbf{J}_R$ is the canonical symplectic matrix
\begin{equation}
   \mathbf{J}_R=
   \begin{bmatrix}
   0 & 1 \\
   -1 & 0 
   \end{bmatrix}\,,\qquad
   \nabla_{\mathbf{X}_\mathcal{B}}=\begin{bmatrix}
   \partial_{p_2} \\ 
   \partial_{\psi_2}
   \end{bmatrix}\,.
\end{equation}
The flow $\dot{\mathbf{X}}_{\mathcal{B}}$ on the shape manifold $\mathcal{B}$ is independent from the flow $\dot{\mathbf{X}}_\mathcal{F}$ along the fiber. Indeed, from \eqref{dynred}
\begin{equation}\label{p2psieq}
	\dot{\psi}_2 =\left(\frac{1}{I_{1}}+\frac{1}{I_{2}}\right)p_{2}-\frac{\mathsf{A}}{I_{1}}\,,\qquad
	\dot{p}_{2} =-\frac{\partial \Pi}{\partial\psi_2}\,.
\end{equation}
On the contrary, $\dot{\mathbf{X}}_\mathcal{F}$ depends on $\dot{\mathbf{X}}_{\mathcal{B}}$ since
\begin{equation}\label{theta1eq}
	\dot{\theta}_1=\frac{\mathsf{A}-p_2}{I_1}\,.
\end{equation} 
In simple words, the flow $\dot{\mathbf{X}}_s$ in the state space $\mathcal{P}$ decouples in a symmetry-free Hamiltonian flow $\dot{\mathbf{X}}_{\mathcal{B}}$ on the shape manifold~$\mathcal{B}$, which induces a rotation drift $\dot{\mathbf{X}}_{\mathcal{F}}$ along the fibers. Physically, the reduced flow on $\mathcal{B}$ is the shape-changing evolution of the connected rotors induced by the internal elastic moments. Such a shape dynamics makes the two rotors to rigidly rotate together by the time-dependent angle $\theta_1$. Thus, given a symmetry-free motion $\mathbf{X}_{\mathcal{B}}$ on $\mathcal{B}$. The full motion in $\mathcal{P}$ follows by shifting $\mathbf{X}_{\mathcal{B}}$ along the fibers by $\theta_1$, that is $\mathbf{X}_s=G_{\theta_1}(\mathbf{X}_{\mathcal{B}})$, as depicted in Fig.~\ref{Fig2}. To evaluate such a rotation drift, from \eqref{1formred}, we define the $1$-form
\begin{equation} \label{1form2}
	\widetilde{\alpha}=\frac{\alpha}{\mathsf{A}}=d\theta_1 +\frac{p_2}{\mathsf{A}} \,d\psi_2\,.
\end{equation}
Then the total rotation drift $\theta_1$ along the fiber follows by integrating the form $d\theta_{1}=\widetilde{\alpha}-\frac{p_{2}}{\mathsf{A}}d\psi_2$, i.e.,
\begin{equation}
	\theta_{1}=\int d\theta_{1}=\int_{0}^{t}\widetilde{\alpha}\,dt-\frac{1}{\mathsf{A}}\int_{\gamma}p_{2}\,d\psi_2\,,
\end{equation}
where $\gamma$ is a closed trajectory of the motion up to time $t$ in the shape manifold $\mathcal{B}$. Thus, 
\begin{equation}
	\theta_{1}=\theta_{\text{dyn}}+\theta_{\text{geom}}\,,  
\end{equation}
where the dynamical and geometric rotation drifts are defined as
\begin{equation} \label{drift}
	 \theta_{\text{dyn}}(t)=\int_{0}^{t}\widetilde{\alpha}\,dt\,,\qquad
	 \theta_{\text{geom}}=-\frac{1}{\mathsf{A}}\int_{\gamma}p_{2}\,d\psi_2\,.
\end{equation}
Here, the dynamical rotation drift $\theta_{\text{dyn}}(t)$ depends on the inertia of the two rotors. Using Eqs.~\eqref{p2psieq}-\eqref{theta1eq} it can written as
\begin{equation}
	\theta_{\text{dyn}}(t)=\frac{2}{\mathsf{A}}\int_{0}^{t}\mathsf{K}(t)\,dt\,,\qquad 
    \mathsf{K}=\frac{1}{2}\left(\frac{p_1^2}{I_1}+\frac{p_2^2}{I_2}\right)\,,
\end{equation}
where $\mathsf{K}$ is the total kinetic energy and $\mathsf{A}$ is the non-zero total angular momentum, which is conserved. If the two rotors are rigidly connected and cannot change their `shape', i.e., $\dot{\psi}_2=0$ (no flow on the base manifold $\mathcal{B}$), then the rotation drift $\theta_{\text{dyn}}$ is simply the manifestation of the inertia of the entire system treated as a whole with angular momentum $\mathsf{A}$ and total kinetic energy $\mathsf{K}$. 

The two rotors can also undergo a change in shape due to the internal elastic moments. As a result, the angle $\psi_2$ varies over time and the flow on $\mathcal{B}$ induces also the geometric rotation drift, which from \eqref{drift} can be written as 
\begin{equation} \label{geom}
	\theta_{\text{geom}}=\int_{S(\gamma)} d\,\widetilde{\alpha}
	=-\frac{1}{\mathsf{A}}\int_{S(\gamma)}d p_{2}\wedge d\psi_2\,.
\end{equation}
Thus, the geometric rotation drift is proportional to the area $S(\gamma)$ enclosed by the trajectory of the motion $\gamma$ in the shape manifold $\mathcal{B}$. 
Such a rotation drift is purely geometric since it does not depend on the time it takes for the two rotors to undergo a cyclic shape change.

One can define an effective moment of inertia $I_{\text{eff}}$ for an equivalent system with the total angular momentum $\mathsf{A}=p_1+p_2$ as
\begin{equation}\label{Ieff}
    \frac{1}{I_{\text{eff}}}=\frac{\dot{\theta}_1}{p_1+p_2}=\frac{\dot{\theta}_{\text{dyn}}+\dot{\theta}_{\text{geom}}}{p_1+p_2}=\frac{2 \mathsf{K}}{\mathsf{A}^2}-\frac{p_2 \,\dot{\psi}_2}{\mathsf{A}^2}\,.
\end{equation}
Using \eqref{p2psieq},
\begin{equation}\label{Ieff}
    \frac{1}{I_{\text{eff}}}=\frac{2 \mathsf{K}}{\mathsf{A}^2}-\frac{p_2^2}{\mathsf{A}^2}+\frac{p_2}{\mathsf{A}\,I_{1}}\,.
\end{equation}
Thus, the shape-changing motion of the two rotors can slow down or speed up the rotation drift. In particular, if the two rotors tend to rotate in opposite directions~($\dot{\psi}_2>0$) the effective moment of inertia increases slowing down the rotation as the angular speed $\dot{\theta_1}$ reduces. On the contrary, if the two rotors tend to rotate in the same direction~($\dot{\psi}_2<0$) the angular speed increases as $I_{\text{eff}}$ reduces. This is the analogue of spinning dancers that can increase their spinning rate by pulling their arms close to their bodies, and to decrease it by letting their arms out. Thus, the geometric phase component in \eqref{Ieff} can be interpreted as the moment of inertia of an  $\textit{added mass}$ in analogy with the fish self-propulsion~\citep{Shapere1987,Shapere1}. See also~\cite{Fedele2023} for a discussion on the effective dynamic mass, including the concept of added mass, in mechanical lattices.

\vskip 0.2in
\begin{figure}[h!]
\centering
\includegraphics[width=0.90\textwidth]{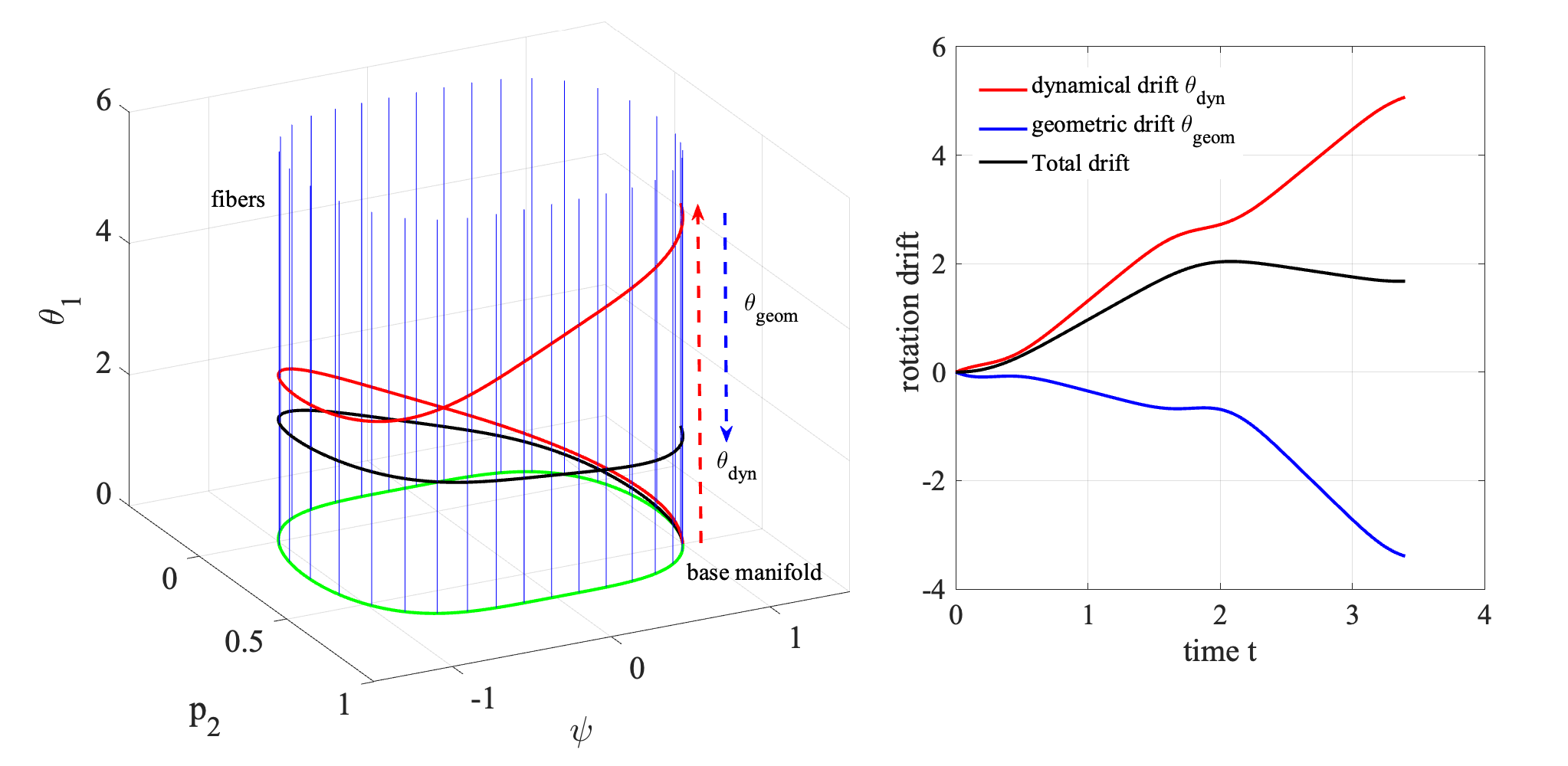}
\caption{(Left) Fiber bundle structure  of the state space $\mathcal{P}=T^{*}Q/\mathsf{A}$ of an elastic double rotor with with parameters $I_1=I_2=1\,\mathrm{mass}\times\mathrm{length}^2$,  $\mathsf{A}=\frac{1}{2}\,\mathrm{mass}\times\mathrm{length}^2\mathrm{time}^{-1}$, and the potential $\Pi(\psi)=\psi^4$. The full path $\mathbf{X}_s(t)$ (black curve) and the reduced path $\mathbf{X}_{\mathcal{B}}(t)$ on the base manifold $\mathcal{B}$ (green curve) are shown. The lifted path $G_{\theta_{\text{dyn}}}(\mathbf{X}_{\mathcal{B}})$ by the dynamical rotation drift $\theta_{\text{dyn}}$ (red curve) does not coincide with with the path $\mathbf{X}_s$ because the total rotation drift $\theta_1=\theta_{\text{dyn}}+\theta_{\text{geom}}$ includes also the geometric component $\theta_{\text{geom}}$. (Right) The total rotation drift $\theta_1$ and the drifts $\theta_{\text{geom}}$ and $\theta_{\text{dyn}}$ as functions of time.}
\label{Fig3}
\end{figure}

As an example, consider the elastic double rotor with parameters $I_1=I_2=1\,\mathrm{mass}\times\mathrm{length}^2$,  $\mathsf{A}=\frac{1}{2}\,\mathrm{mass}\times\mathrm{length}^2\mathrm{time}^{-1}$, and the potential $\Pi(\psi)=\psi^4$. The total angular momentum is assumed to be positive and the initial conditions are chosen so that both rotors have positive (counter-clockwise) rotational speed. The left panel of Fig.~\ref{Fig3} depicts the fiber bundle structure  of the state space $\mathcal{P}=T^{*}Q/\mathsf{A}$ of the elastic double rotor, the full path $\mathbf{X}_s(t)$~(black curve) and the reduced path $\mathbf{X}_{\mathcal{B}}$~(green curve) on the base manifold $\mathcal{B}$ are shown. The lifted path $G_{\theta_{\text{dyn}}}(\mathbf{X}_s)$ by the positive dynamical rotation drift~$\theta_{\text{dyn}}$ (red curve) does not coincide with the path $\mathbf{X}_s$ (black curve). This is because the total rotational drift $\theta_1=\theta_{\text{dyn}}+\theta_{\text{geom}}$ includes also a negative geometric component $\theta_{\text{geom}}$, as is seen in the right panel of the same figure. The positive dynamical rotation drift is induced by the inertia of the system that has a positive angular momentum. However, the two rotors undergo changes in shape inducing a clockwise rotation that balances the counter-clockwise dynamical drift rotation. 

\begin{remark}\label{Sign-Momentum}
Given the total angular momentum $\mathsf{A}$, the associated dynamical variables $\{\theta_1(t),\psi_2(t),p_2(t)\}$ satisfy Eqs.~\eqref{p2psieq} and~\eqref{theta1eq}. Let $\{\bar{\theta}_1(t),\bar{\psi}_2(t),\bar{p}_2(t)\}$ be the dynamical variables that correspond to $-\mathsf{A}$, where $\bar{\psi}_2=\bar{\theta}_2-\bar{\theta}_1$. Then, these satisfy  
\begin{equation}
	\dot{\bar{\theta}}_1=\frac{-\mathsf{A}-\bar{p}_2}{I_1}	\,,\quad 
    \dot{\bar{\psi}}_2 =\left(\frac{1}{I_{1}}+\frac{1}{I_{2}}\right)\bar{p}_{2}
	+\frac{\mathsf{A}}{I_{1}}\,,\qquad
	\dot{\bar{p}}_{2} =-\frac{\partial \Pi}{\partial\bar{\psi}_2}\,.
\end{equation}
Assume the initial conditions $\bar{\theta}_1(0)=-\theta_1(0)$, $\bar{\psi}_2(0)=-\psi_2(0)$, and $\bar{p}_2(0)=-p_2(0)$. Then, from Eqs.~\eqref{p2psieq} and~\eqref{theta1eq} we have 
\begin{equation}
	\left\{\bar{\theta}_1(t),\bar{\psi}_2(t),\bar{p}_2(t)\right\}
	=\left\{-\theta_1(t),-\psi_2(t),-p_2(t)\right\}\,,
\end{equation}
is a solution of the above system of first-order ODEs. 
The associated symplectic form follows from \eqref{1formred} as
\begin{equation}
	\bar{\alpha}=-\mathsf{A} \,d\bar{\theta}_1 + \bar{p}_2 \,d\bar{\psi}_2\,,
	\qquad d\bar{\alpha}= d\bar{p}_2\wedge d\bar{\psi}_2\,. 
\end{equation}
Moreover, 
\begin{equation} 
	\widetilde{\bar{\alpha}}=\frac{\bar{\alpha}}{-\mathsf{A}}=-\widetilde{\alpha}\,.
\end{equation}
Therefore
\begin{equation} 
	 \bar{\theta}_{\text{dyn}}(t)=\int_{0}^{t}\widetilde{\bar{\alpha}}\,dt
	 =- \theta_{\text{dyn}}(t)\,,\qquad
	 \bar{\theta}_{\text{geom}}=\frac{1}{\mathsf{A}}\int_{\gamma}\bar{p}_{2}\,d\bar{\psi}_2
	 =- \theta_{\text{geom}}\,.
\end{equation}
Thus, when changing the sign of the angular momentum (and the initial conditions), both the dynamic and geometric phases change sign.
\textcolor{black}{This implies that one has the freedom to set a clockwise angular momentum as either positive or negative by simply flipping the frame, or coordinate chart. Similarly, the same thing can be done for counterclockwise angular momenta.
For example, consider the rotor system with a clockwise angular momentum $\mathsf{A}>0$ defined as positive in the frame $\{\theta_1,p_2,\psi_2\}$. In the flipped frame $\{-\theta_1,-p_2,-\psi_2\}$, the angular momentum bears the opposite sign, $\mathsf{A}<0$, but it still preserves its clockwise orientation.
The orientation of the flipped frame changes because the Jacobian determinant of the transformation $\{\theta_1 \rightarrow -\theta_1,p_2 \rightarrow -p_2,\psi_2 \rightarrow -\psi_2\}$ is negative reflecting the change in sign of $\mathsf{A}$. This suggests that one can define a given angular momentum to be either positive or negative without lose of generality.}
\end{remark}

\subsection{Curvature and intrinsic metric of the shape manifold}

One can interpret the geometric drift as the curvature of the shape manifold $\mathcal{B}$ equipped with a specific metric. As a matter of fact, drawing on Cartan's structural equations the $2$-form $d\,\widetilde{\alpha}$ in~\eqref{geom} can be interpreted as the curvature form of a connection on $\mathcal{B}$. We further require that the symplectic form be compatible with the volume 2-form $\mathsf{vol}_{\mathbf{G}}$ of the metric $\mathbf{G}$ as is shown next.

The geometric drift $\theta_{\text{geom}}$ given in \eqref{geom} is associated to the symplectic $1$-form
\begin{equation}\label{1form2}
	\widetilde{\alpha}=-\frac{p_2}{\mathsf{A}}\, d\psi_2\,,
\end{equation}
since $\theta_{\text{geom}}=\int d\,\widetilde{\alpha}$.
The $1$-form $\widetilde{\alpha}$ can be interpreted as the connection $1$-form $\omega^1{}_2$ of a $2$-manifold represented by the coordinate charts $\{X^1,X^2\}=\{p_2,\psi_2\}$ and with the metric
\begin{equation}\label{2dmetric}
	ds^{2}=\epsilon_1\,\mathsf{G}_1\,dp_2^2+\epsilon_2\,\mathsf{G}_2\,d\psi_2^2\,. 
\end{equation}
We want to find the metric coefficients $\mathsf{G}_1$ and $\mathsf{G}_2$ so that\footnote[1]{We can add to $\omega^1{}_2$ an arbitrary closed $1$-form $\xi$, i.e., $d \xi=0$. This form can be neglected because it does not contribute to the geometric phase as its integral over any closed curve vanishes. Thus, the freedom to add an arbitrary closed form is physically inconsequential.}
\begin{equation}\label{1form3}
    \omega^1{}_2=\widetilde{\alpha}\,.
\end{equation}
From \eqref{connectionforms} we then have
\begin{equation} 
	\frac{\mathsf{G}_{1,2}}{2\sqrt{\mathsf{G}_{1}\mathsf{G}_{2}}}\,dp_2
	-\epsilon_1\epsilon_2\,\frac{\mathsf{G}_{2,1}}{2\sqrt{\mathsf{G}_{1}\mathsf{G}_{2}}}\,d\psi_2
	=-\frac{p_2}{\mathsf{A}} \,d\psi_2\,.
\end{equation}
This implies that
\begin{equation}\label{G1G2eq}
	\frac{\mathsf{G}_{1,\psi_2}}{2\sqrt{\mathsf{G}_{1}\mathsf{G}_{2}}}=0\,,\qquad
	\epsilon_1\epsilon_2\,\frac{\mathsf{G}_{2,p_2}}{2\sqrt{\mathsf{G}_{1}\mathsf{G}_{2}}}=\frac{p_2}{\mathsf{A}}\,.
\end{equation}
The first equation implies that $\mathsf{G}_{1,\psi_2}=0$, and hence $\mathsf{G}_1=\mathsf{G}_1(p_2)$. 
The Gaussian curvature is calculated from \eqref{Gaussiancurvature} as
\begin{equation}\label{KG}
	K=-\frac{1}{\mathsf{A}\sqrt{\mathsf{G}_1\mathsf{G}_2}}\,.
\end{equation}

\textcolor{black}{Thus, the curvature depends on the sign of $\mathsf{A}$. This is a consequence of matching the symplectic and curvature forms in \eqref{1form3}. In Remark~\ref{Sign-Momentum} it was shown that by flipping the coordinate chart, one can consistently define the rotation sign of the angular momentum $\mathsf{A}$ to be either positive or negative, e.g., counterclockwise or clockwise, respectively, or viceversa. Consequently, the base manifold of the elastic rotor system can be endowed with two distinct metrics, depending on the convention used to define the sign of the total angular momentum.} 

Eqs.~\eqref{G1G2eq} also imply that the symplectic $2$-form $d\,\widetilde{\alpha}$ is equal to the curvature $2$-form $d\omega^1{}_2$, that is $d\,\widetilde{\alpha}=d\omega^1{}_2$, or explicitly 
\begin{equation}
    d\,\widetilde{\alpha}=-\frac{1}{\mathsf{A}} \,dp_2\wedge d\psi_2 = K\sqrt{\mathsf{G}_1\mathsf{G}_2}\,dp_2\wedge d\psi_2\,.
\end{equation}
We now further require that the symplectic $2$-form $d\,\widetilde{\alpha}$ is compatible with the (pseudo) Riemannian volume~(area) $2$-form $\mathsf{vol}_{\mathbf{G}}=\sqrt{\mathsf{G}_1\mathsf{G}_2} \,dp_2\wedge d\psi_2$ in the sense that the absolute value of the geometric rotational drift over a closed trajectory $\gamma$ in the shape manifold is equal to the volume (area) of the region $S(\gamma)$ it encloses using $\mathsf{vol}_{\mathbf{G}}$, that is,
\begin{equation}
    \frac{1}{|\mathsf{A}|} \,dp_2\wedge d\psi_2 = \sqrt{\mathsf{G}_1\mathsf{G}_2}\,dp_2\wedge d\psi_2\,,
\end{equation}
which is equivalent to
\begin{equation}\label{G1G2}
    \sqrt{\mathsf{G}_1\mathsf{G}_2}=\frac{1}{|\mathsf{A}|}\,,
\end{equation}
or $\mathsf{A}^2\,\mathsf{G}_1\,\mathsf{G}_2=1$.
This, in particular, implies that $G_2=G_2(p_2)$.
We can now solve for $G_2$. Substituting \eqref{G1G2} into~\eqref{G1G2eq}$_2$ yields
\begin{equation}\label{Geq}
	\mathsf{G}_{2,p_2}=\epsilon_1\epsilon_2\,\frac{2\,p_2}{\mathsf{A}|\mathsf{A}|}\,,
\end{equation}
and hence
\begin{equation}\label{Geq}
	\mathsf{G}_2=\epsilon_1\epsilon_2\,\frac{p_2^2+C_2}{\mathsf{A}|\mathsf{A}|}\,,
\end{equation}
where $C_2$ is an arbitrary constant, and $G_1$ follows from \eqref{G1G2}. Since we must have $G_2\geq 0$, we set $C_2=\mu^2$, $\mu\in\mathbb{R}$, and $\mathrm{sgn}(\epsilon_1\epsilon_2)=\mathrm{sgn}(\mathsf{A})$. We choose $\epsilon_1=\mathrm{sgn}(\mathsf{A})$ and $\epsilon_2=1$\footnote{Another choice would be $\epsilon_1=1$ and $\epsilon_2=\mathrm{sgn}(\mathsf{A})$, which gives the following metric
\begin{equation}
	\mathbf{G}^*=\frac{1}{p_2^2+\mu^2}\,dp_2\otimes d p_2
	+\mathrm{sgn}(\mathsf{A})\frac{p_2^2+\mu^2}{\mathsf{A}^2}\,d\psi_2\otimes d\psi_2\,,
\end{equation}
and $\mathbf{G}^*=\mathrm{sgn}\,(\mathsf{A})\mathbf{G}$. The two metric yield the same curvature $K$. They are identical for $\mathsf{A}>0$ and Riemannian in character. For $\mathsf{A}<0$, we have $\mathbf{G}^*=-\mathbf{G}$ and the two metrics are pseudo-Riemannian. 
} so that
\begin{equation}
	\mathsf{G}_1=\frac{1}{p_2^2+\mu^2}\,,\qquad 
    \mathsf{G}_2=\frac{p_2^2+\mu^2}{\mathsf{A}^2}\,,
\end{equation}
and the family of metrics~\eqref{2dmetric} is simplified to read
\begin{equation}\label{metric2}
	\mathbf{G}=\frac{\mathrm{sgn}(\mathsf{A})}{p_2^2+\mu^2}\,dp_2\otimes d p_2+\frac{p_2^2+\mu^2}{\mathsf{A}^2}\,d\psi_2\otimes d\psi_2\,. 
\end{equation}
From \eqref{KG} and~\eqref{G1G2}, one obtains the corresponding Gaussian curvature $K=-\mathrm{sgn}(\mathsf{A})$, and the Ricci scalar $R=2 K=-2\,\mathrm{sign}(\mathsf{A})$. \textcolor{black}{ The metric $\mathbf{G}$ and the corresponding curvatures depend on the sign of $\mathsf{A}$. Notably, as highlighted in Remark~\ref{Sign-Momentum}, one has the freedom to define the sign of the angular momentum to be either positive or negative. This implies the existence of two distinct metrics, mirroring the convention used to define the rotation sign.
In the subsequent sections, we demonstrate that choosing $\mathsf{A}<0$ endows the shape manifold $\mathcal{B}$ with the pseudo-Riemannian structure of an Einstein metric~\citep{Besse:Besse1987} of the sectional plane of an expanding $4$D~spacetime with positive curvature, equipped with the Robertson-Walker metric~\citep{GRAVITATION,Carroll2003spacetime}. Conversely, opting for $\mathsf{A}>0$ the shape manifold is endowed with the structure of the hyperbolic plane with negative curvature. For both cases, the geometric phase is evaluated by the same $2$-form, derived from the sectional curvature form of~$\mathcal{B}$ in~\eqref{geom}. The two metrics are compatible with the geometric phase, except for its sign, mirroring the sign convention used. Moreover, the two metrics have different curvatures and cannot be isometric. 
}

\begin{remark}[Metric Uniqueness]\label{2rotoruniquemetric}
\textcolor{black}{The metric depends on the sign of $\mathsf{A}$ because we matched the symplectic $1$-form $\widetilde{\alpha}=-\frac{p_2}{\mathsf{A}}\, d\psi_2$ in \eqref{1form2} with the curvature form of the base manifold~$\mathcal{B}$ in \eqref{1form3}. In doing so, the intent is to have curvature equal to the geometric phase in \eqref{geom}. This depends on the sign of~$\mathsf{A}$ and curvature inherits it.
Alternatively, a unique metric can be defined by matching the symplectic form~$\beta=-\mathsf{A}\,\widetilde{\alpha}=p_2\, d\psi_2$ of the reduced dynamics on~$\mathcal{B}$ with the curvature form in \eqref{1form3}. In this case the curvature is set to be equal to the area spanned by the Hamiltonian flow on the base manifold. As a result, the geometric phase is proportional to curvature, with constant of proportionality  $-\frac{1}{\mathsf{A}}$, see \eqref{drift}. Such a matching equips $\mathcal{B}$ with the following pseudo-Riemannian metric
\begin{equation}\label{metric4}
	\mathbf{G}=-\frac{1}{p_2^2+\mu^2}\,dp_2\otimes d p_2+(p_2^2+\mu^2)\,d\psi_2\otimes d\psi_2\,, 
\end{equation}
which is an Einstein metric~\citep{Besse:Besse1987}. In particular, this is a disguised metric of the $2$D~section of a $4$D Robertson-Walker expanding spacetime universe for any real number $\mu$, as hereafter shown.}
\end{remark}

\begin{remark}
\citet{HernandezGarduno2018} studied the geometric phases of three inviscid point vortices and also found that the curvature of the associated shape manifold depends on the sign of a parameter related to the strengths and circulations of the three vortices. In particular, their shape manifold is a sphere, for example if the three vortices spin in the same direction forming a vortex cluster. It is instead a hyperbolic plane, for example when one of the vortex spins opposite to the other two vortices. Thus, the change in character of the manifold signals different vortex interactions~(see also~\cite{leapfrogvortexShashikanthMarsden2003}). We note that in their system each vortex interacts with the other two, allowing for non-trivial dynamical configurations. 
\end{remark}

\begin{remark}[Physical significance of the metric]
The metric~\eqref{metric2} defined on the shape, or base manifold $\mathcal{B}$ characterizes the kinematically admissible shape deformations of the elastic double rotor. An orbit on $\mathcal{B}$ is a succession of  infinitesimal changes in the shape of the elastic double rotor from an initial configuration to another. If the elastic double rotor returns to its initial shape, the orbit is closed and the area~(or curvature) spanned by it measures the induced rotation drift. Any curve, or orbit on the base manifold is a kinematically admissible shape evolution, i.e., a sequence of changing shapes. The orbit is also dynamically admissible if it is consistent with the Hamiltonian flow~\eqref{p2psieq}. 
The metric allows quantifying the similarity of a shape $S_1$ to another shape $S_2$, by measuring the intrinsic distance between the corresponding points on the shape manifold~$\mathcal{B}$. Other non-intrinsic distances would be misleading as they do not account for the curvature, or induced geometric drift. The important point is that different shapes must be compared using the same metric chosen based on the sign of the angular momentum $A$.  
\textcolor{black}{In the following sections we will show for~$\mathsf{A}>0$ the distance between two shapes with different momenta $p_2$ appear red-shifted and their distance is larger than the corresponding Euclidean distance. Similarly, for $\mathsf{A}<0$ the two shapes appear distant in the hyperbolic plane in comparison to what one would observe in the Euclidean plane. So, the two metrics qualitatively describe the intrinsic differences in shapes, which is misled as shorter through the Euclidean lens.} 
\end{remark}

\subsection{Geodesics of the metric}

\subsubsection{Negative angular momentum: $2$D Robertson-Walker spacetime}

\begin{figure}[h!]
\centering
\includegraphics[width=0.55\textwidth]{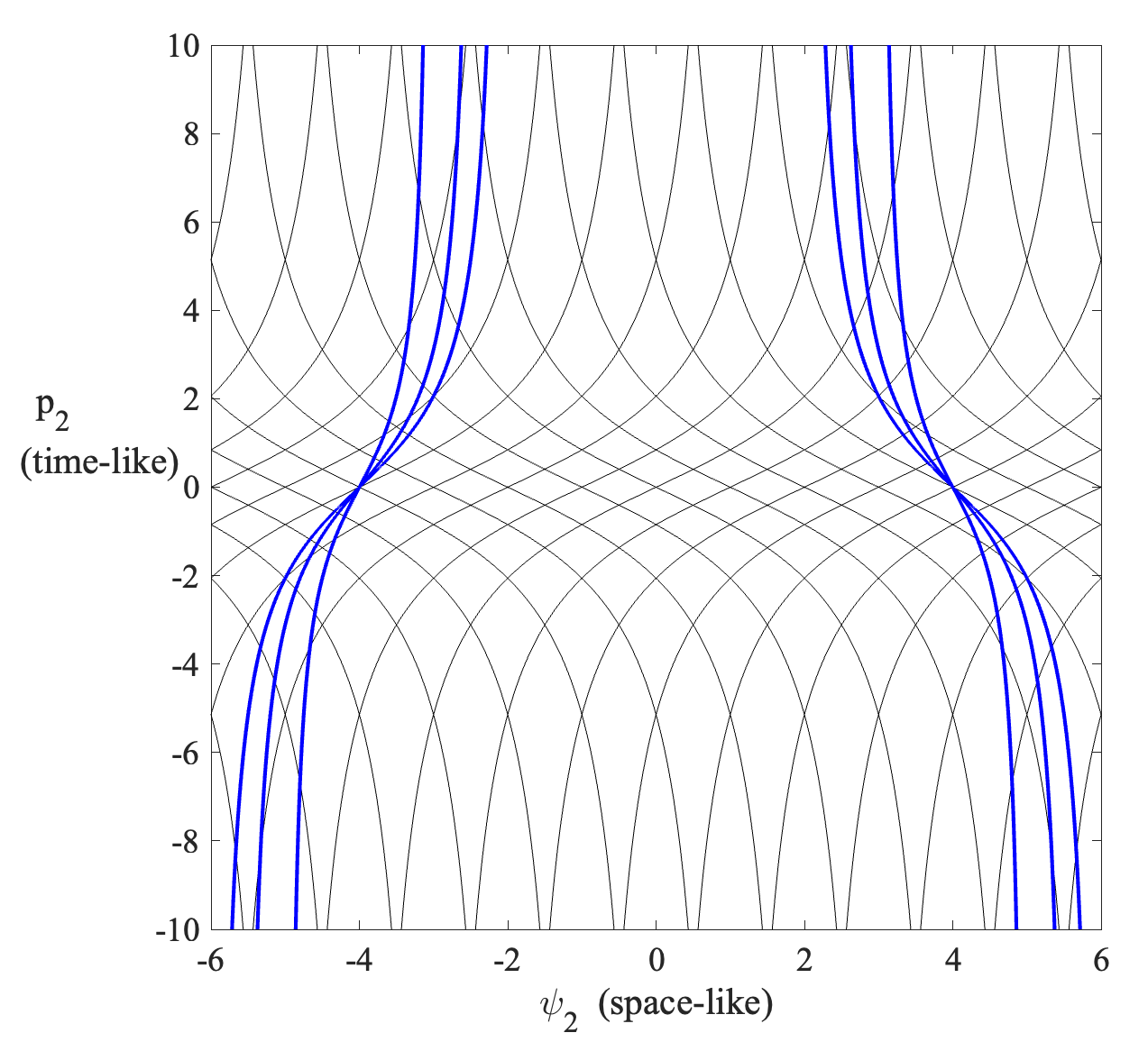}
\vskip 0.1in
\caption{Geodesics of the pseudo Riemannian metric~$-\frac{1}{p_2^2+\mu^2}\,dp_2\otimes d p_2+\frac{p_2^2+\mu^2}{\mathsf{A}^2}\,d\psi_2\otimes d\psi_2$. Null-geodesics are the thin black curves and geodesics are the bold blue curves.} 
\label{Fig4}
\end{figure}

\textcolor{black}{Choosing $\mathsf{A}<0$, the shape manifold has positive Gaussian curvature $K=1$} and the metric in \eqref{metric2} is pseudo-Riemannian with $p_2$ as a time-like coordinate and $\psi_2$ as space-like, 
\begin{equation}\label{pseudometric}
    ds^2=-\frac{1}{p_2^2+\mu^2}\,(dp_2)^2+\frac{p_2^2+\mu^2}{\mathsf{A}^2}\,(d\psi_2)^2\,.
\end{equation}
The geodesic equations follow by minimizing the action $\int ds^2=\int L \,d\lambda$ with Lagrangian density 
\begin{equation}\label{Lagrlambda}
	L(p_2(\lambda),\psi_2(\lambda))=-\frac{(p^{\prime}_2)^2}{p_2^2+\mu^2} + \frac{p_2^2+\mu^2}{\mathsf{A}^2}(\psi^{\prime}_2)^2\,,
\end{equation}
where $f^{\prime}$ denotes derivative with respect to $\lambda$, which parameterizes the geodesics. Trivial geodesics are the straight lines $\psi_2=\mathrm{const}$ and $p_2=\mathrm{const}$ for which $L$ is stationary. A family of non-trivial geodesics can be easily found by choosing the parametrization $\lambda=p_2$. 
Then, 
\begin{equation}\label{Lagr}
	L(\psi_2(p_2))=-\frac{1}{p_2^2+\mu^2} + \frac{p_2^2+\mu^2}{\mathsf{A}^2}\left(\frac{d\psi_2}{dp_2}\right)^2\,.
\end{equation}
Variational differentiation gives
\begin{equation}
   \frac{d}{d p_2}\left(\frac{p_2^2+\mu^2}{\mathsf{A}^2}\,\frac{d\psi_2}{dp_2}\right)=0\,,
\end{equation}
from which
\begin{equation}\label{geodequa45}
   \frac{d\psi_2}{dp_2}=\frac{\mathsf{A}^2 C_1}{p_2^2+\mu^2}\,,
\end{equation}
where $C_1$ is an arbitrary constant. Integration yields
\begin{equation}\label{geodesic}
  \psi_2=\frac{\mathsf{A}^2 C_1}{\mu}\,\tan^{-1}\left(\frac{p_2}{\mu}\right)+C_2\,,
\end{equation}
where $C_2$ is another arbitrary constant, which together with $C_1$ parameterize the family of geodesics. The Lagrangian density in \eqref{Lagr} simplifies to read
\begin{equation}
    L=\frac{-1+\mathsf{A}^2 C_1^2}{p_2^2+\mu^2}\,,
\end{equation}
and the null-geodesics are given by \eqref{geodesic} with $C_1=1/\mathsf{A}$ since $L=0$. Fig.~\ref{Fig4} depicts the null-geodesics (thin black curves) and a few geodesics (bold blue curves) for the metric with $\mathsf{A}=-5,\mu=2$.

\begin{remark}
Drawing on General Relativity~\citep{GRAVITATION,Carroll2003spacetime}, $p_2$ is a time-like coordinate and $\psi_2$ is space-like, and the metric $\mathbf{G}$ represents the analogue of a space-time where null-geodesics (thin curves of Fig.~\ref{Fig4}) are the trajectories of massless light photons. The associated light cones tend to close up as $p_2 \rightarrow \pm \infty$ and light slows down. In the same figure, the depicted geodesics (bold curves) are always inside the light cones they intersect along their path. Thus, they are the `time-like' trajectories of a massive particle traveling at a speed less than the speed of light. Moreover, geodesics tend to converge in as an indication of the positive Gaussian curvature. 
\end{remark}

\begin{remark}
The metric~\eqref{pseudometric} describes the analogue of a disguised sectional plane of a $4$D~spacetime of an expanding universe~\citep{GRAVITATION,Carroll2003spacetime}. As a matter of fact, for $\mu\neq 0$, one can define the following new coordinate chart 
\begin{equation}\label{coordchange1}
   t=\tanh^{-1}\left[\frac{p_2}{\sqrt{p_2^2+\mu^2}}\right],\qquad x=\psi_2\,. 
\end{equation}
Then one has $dt=d p_2/\sqrt{p_2^2+\mu^2}$, $ dx=d\psi_2$, and the metric~\eqref{pseudometric} transforms to
\begin{equation}\label{RW}
    ds^2=-dt^2 + \frac{\mu^2 (\cosh{t})^2}{\mathsf{A}^2}\, dx^2\,,
\end{equation}
\textcolor{black}{which is the induced metric on the $2$D~section $(t,x)$ of the $4$-dimensional Robertson-Walker (RW) spacetime in General Relativity~\citep{GRAVITATION,Carroll2003spacetime}. The RW~metric
\begin{equation}
   ds^2_{RW}=-dt^2 + \frac{\mu^2 (\cosh{t})^2}{\mathsf{A}^2}\,(dx^2+dy^2+dz^2)\,, 
\end{equation}
 describes an expanding universe with scale factor $a(t)=\cosh{t}$ and Hubble constant $H=\frac{\dot{a}}{a}=\tanh t$~\citep{Carroll2003spacetime}.} For $\mu=0$, the new coordinates are
\begin{equation}\label{coordchange2}
   t=\mathrm{e}^{p_2},\qquad x=\psi_2\,, 
\end{equation}
where $dt=\frac{d p_2}{p_2}$ and $ dx=d\psi_2$, and the metric transforms to
\begin{equation}
    ds^2=-dt^2 + \frac{\mathrm{e}^{2t}}{\mathsf{A}^2} \,dx^2\,,
\end{equation}
which is still the induced metric on the section $(x,t)$ of the Robertson-Walker spacetime with the scale factor $a(t)=\mathrm{e}^t$ and Hubble constant $H=\frac{\dot{a}}{a}=1$~\citep{Carroll2003spacetime}. \textcolor{black}{In the following, the  metrics~\eqref{pseudometric},\eqref{RW} will be referred to as the metrics of a $2$D Robertson-Walker spacetime universe.}
\end{remark}

\begin{remark}
The analogy of the shape manifold~$\mathcal{B}$ being like an expanding universe implies that  a point on $\mathcal{B}$, or shape $S_1$, appears `red-shifted' by another point, or shape $S_2$, as the momentum $p_2$ (time-like coordinate) increases. Thus, the low-momentum shapes with small geometric rotation drift are far apart from the high-momentum shapes with large geometric drift.  
So the analogy with the expanding universe implies that different shapes, or points,  can be very far away from each other on the shape manifold and correspond to very different geometric rotation drifts. The extrinsic Euclidean metric would give a smaller distance between the two points misleading them as similar shapes. 
\end{remark}

\subsubsection{Positive angular momentum: The hyperbolic plane $\mathbb{H}^2$}

\textcolor{black}{Choosing $\mathsf{A}>0$}, the shape manifold has negative Gaussian curvature $K=-1$ and the metric in \eqref{metric2} is Riemannian 
\begin{equation}\label{Riemannmetric}
    ds^2=\frac{1}{p_2^2+\mu^2}\,(dp_2)^2+\frac{p_2^2+\mu^2}{\mathsf{A}^2}\,(d\psi_2)^2\,.
\end{equation}
To reveal the nature of the geodesics, we can still use the coordinate transformations~\eqref{coordchange1},~\eqref{coordchange2}. As an example, for $\mu=0$ the metric~\eqref{Riemannmetric} transforms to $ds^2=dt^2 + R^2(t) \,dx^2$, where $R(t)=\mathrm{e}^{2t}/\mathsf{A}^2$. This is a disguised metric of the hyperbolic plane as the change of coordinates $\tilde{x}=x/\sqrt{\mathsf{A}}$ and $\tilde{y}=\mathsf{A}\exp(-t)$ transforms it to $ds^2=\frac{d\tilde{x}^2+d\tilde{y}^2}{\tilde{y}^2}$.

\begin{remark}
A point on $\mathcal{B}$, or shape $S_1$, appears far away from another point, or shape $S_2$, as the momentum $p_2$ reduces because of the hyperbolic character of the metric. Thus, the low-momentum shapes with small geometric rotation drift are far apart from the high-momentum shapes with large geometric drift. If one uses the Euclidean metric instead, the two points would appear closer than they are. The Euclidean metric is misleading in the sense that far away shapes appear as similar shapes when looking at them through a Euclidean lens. 
\end{remark}

\subsubsection{The set of all geodesics}

More generally, let us assume that the geodesics of the metric in \eqref{metric2} are parameterized by $\lambda$. Minimizing the action $\int ds^2=\int L \,d\lambda$ with Lagrangian density 
\begin{equation}\label{Lagrlambda666}
	L(p_2(\lambda),\psi_2(\lambda))=\mathrm{sgn}(\mathsf{A})\frac{(p^{\prime}_2)^2}{p_2^2+\mu^2} + \frac{p_2^2+\mu^2}{\mathsf{A}^2}(\psi^{\prime}_2)^2\,,
\end{equation}
gives
\begin{equation}\label{geodeq45a}
   \left(\frac{p_2^{\prime}}{p_2^2+\mu^2}\right)^{\prime}+\mathrm{sgn}(\mathsf{A})\frac{p_2}{ \mathsf{A}^2}\,{\psi^{\prime}_2}^2-\frac{p_2 p_2^{\prime}}{p_2^2+\mu^2}=0\,,\qquad \psi^{\prime}_2=\frac{\mathsf{A}^2 c_1}{p_2^2+\mu^2}\,,
\end{equation}
where $f^{\prime}$ denotes derivative with respect to $\lambda$ and $c_1$ is a constant. Then, the first equation for $p_2$ can be written as
\begin{equation}
 p_2^{\prime\prime}=p_2\frac{-\mathrm{sgn}(\mathsf{A})\mathsf{A}^2 c_1^2 + p_2^{\prime}}{p_2^2+\mu^2}\,. 
\end{equation}
This ODE can be solved for by using the substitution $p_2^{\prime}=F(p_2)$. Notice that $p_2^{\prime \prime}=\frac{dF}{d p_2} p_2^{\prime}= F\,\frac{dF}{d p_2}$ and 
\begin{equation}
    \frac{dF}{d p_2} F=p_2\frac{-\mathrm{sgn}(\mathsf{A})\mathsf{A}^2 c_1^2 + F}{p_2^2+\mu^2}\,,
\end{equation}
which can be easily integrated to solve for $F$:
\begin{equation}
    p_2^{\prime}=F(p_2)=\pm \sqrt{  c_2\,(p_2^2+\mu^2)+\mathrm{sgn}(\mathsf{A})\mathsf{A}^2 c_1^2 }\,,
\end{equation}
where $c_2$ is another constant. Thus, the geodesic equations~\eqref{geodeq45a} are reduced to the following first order system
\begin{equation}
    p_2^{\prime}=\pm \sqrt{ c_2\,(p_2^2+\mu^2)+\mathrm{sgn}(\mathsf{A})\mathsf{A}^2 c_1^2}\,,\qquad\quad \psi^{\prime}_2=\frac{\mathsf{A}^2 c_1}{p_2^2+\mu^2}\,. 
\end{equation}
Integrating the first equation yields
\begin{equation}
    -\frac{1}{\sqrt{c_2}}\,\log\left(-p_2\sqrt{c_2}  +\sqrt{c_2\,(p_2^2+\mu^2)+\mathrm{sgn}(\mathsf{A})\mathsf{A}^2 c_1^2} \right)
 = \pm \lambda + \lambda_0\,,   
\end{equation}
which is valid in the range of values of $\lambda$ for which the argument under the square root is non-negative, $\lambda_0$ is a constant, and $c_2\ge 0$. Thus, the geodesics are parameterized by 
\begin{equation}
 p_2(\lambda)=\frac{\mathrm{e}^{-\sqrt{c_2} (\pm \lambda + \lambda_0)} \left[-1 + \left(c_2 \mu^2+\mathrm{sgn}(\mathsf{A})\,\mathsf{A}^2 c_1^2\right) \mathrm{e}^{2 \sqrt{c_2} (\pm \lambda + \lambda_0)}\right]}{2\sqrt{c_2}}\,,
\end{equation}
and
\begin{equation}
    \psi_{2}(\lambda)=\begin{cases}
\begin{array}{c}
\frac{\mathsf{A}c_{1}}{\mu}\tanh^{-1}\left[\frac{2\mathsf{A}\,\mu\,c_{1}\sqrt{c_{2}}\,\mathrm{e}^{2\sqrt{c_{2}}(\pm\lambda+\lambda_{0})}}{1+\left(\mathsf{A}^{2}\,c_{1}^{2}+c_{2}\,\mu^{2}\right)\mathrm{e}^{2\sqrt{c_{2}}(\pm\lambda+\lambda_{0})}}\right],\qquad\qquad\qquad\qquad\qquad\qquad\qquad\qquad\qquad\qquad \mathsf{A}<0\,,\\
\\
\frac{\mathsf{A}\,c_{1}}{\mu}\Bigg\{\tanh^{-1}\left[\frac{A\,c_{1}+\left(A^{2}c_{1}^{2}+c_{2}\,\mu^{2}\right)\mathrm{e}^{\sqrt{c_{2}}(\pm\lambda+\lambda_{0})}}{\sqrt{c_{2}}\,\mu}\right]-\tanh^{-1}\left[\frac{Ac_{1}-\left(A^{2}c_{1}^{2}+c_{2}\,\mu^{2}\right)\mathrm{e}^{\sqrt{c_{2}}(\pm\lambda+\lambda_{0})}}{\sqrt{c_{2}}\mu}\right]\Bigg\}\,,\quad\mathsf{A}>0\,.
\end{array}\end{cases}
\end{equation}

\begin{figure}[h!]
\vskip 0.2in
\centering
\includegraphics[width=0.6\textwidth]{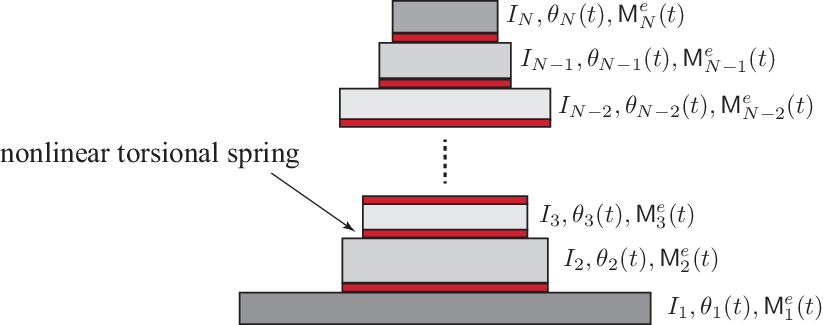}
\vskip 0.1in
\caption{The side view of an elastic $N$-rotor with $N-1$ nonlinear springs. The applied time-dependent moments are self equilibrated, i.e., $\sum_{j=1}^{N}\mathsf{M}^e_j(t)=0$.}\label{N-pendulum}
\label{FigN}
\end{figure}

\section{Dynamics of (free) nonlinear elastic $N$-rotors}

We next generalize the elastic double rotor system described above to an elastic $N$-rotor with $N$ rotors with mass moments of inertia $(I_1,I_2, ...I_N)$. The Lagrangian coordinates $\theta_j$ are the angular positions of the rigid rotors as depicted in Fig.~\ref{N-pendulum}. For the specific problem we consider the action of $N-1$ nonlinear springs on the rotors as depicted in Figure~\ref{N-pendulum}. The associated potential depends on the angle differences of adjacent rotors, i.e,
\begin{equation}
    \Pi\left(\theta_2-\theta_1,\theta_3-\theta_2,...,\theta_{N}-\theta_{N-1}\right)=\Pi_2(\theta_2-\theta_1)+\hdots+\Pi_{N}(\theta_{N}-\theta_{N-1})\,,
\end{equation}
and describes internal conservative moments $M_j=-\partial_{\theta_j}\Pi$ acting on the rotors, which are in equilibrium, that is
\begin{equation}\label{balanceN}
	M_1+M_2+\cdots+M_N=-\sum_{j=1}^{N}\frac{\partial \Pi}{\partial\theta_{j}}=0\,. 
\end{equation}
The associated Lagrangian is written as
\begin{equation} \label{LagrN}
    \mathcal{L}=\sum_{n=1}^N\frac{1}{2}I_{n}\,\dot{\theta}_{n}^{2}- \Pi\left(\theta_2-\theta_1,\theta_3-\theta_2,...,\theta_{N}-\theta_{N-1}\right)\,.
\end{equation}
Minimizing the action $\int \mathcal{L} dt $ yields the following dynamical equations
\begin{equation} \label{LagrangeeqsN}
 	\frac{d}{dt}\left(\frac{\partial\mathcal{L}}{\partial\dot{\theta}_j}\right)
	-\frac{\partial\mathcal{L}}{\partial\theta_{j}}=I_j\ddot{\theta}_j+\frac{\partial \Pi}{\partial\theta_j}=0\,,
	\qquad j=1,\cdots N\,.
\end{equation}
From~\eqref{balanceN} potential moments are in equilibrium and summing up equations~\eqref{LagrangeeqsN} yields 
\begin{equation} 
	\frac{d}{dt}\left(\sum_{j=1}^N I_j\dot{\theta}_j(t)\right)=0\,.
\end{equation}
Thus, the total angular momentum of the elastic $N$-rotor
\begin{equation}\label{angulmomemN}
    I_{1}\dot{\theta}_{1}(t)+I_2\dot{\theta}_2(t)+\cdots+I_N\dot{\theta_N}(t)=\mathsf{A} \,,
\end{equation}
 is conserved over time and $\mathsf{A}=I_1\dot{\theta_1}(0)+\cdots+I_N\dot{\theta_N}(0)$ is the initial momentum imparted by the angular velocities $\dot{\theta_j}$ at time $t=0$. We  can associate a Hamiltonian system on the cotangent space $T^*Q$ of the configuration space $Q=\mathbb{T}^N$, that is the $N$-torus with a coordinate chart $\{\theta_1,\theta_2,\cdots,\theta_N\}$. The conjugate momenta of the angles $\theta_j$ are $p_j=\partial_{\dot{\theta}_j}\mathcal{L}=I_j\dot{\theta}_j$. Thus, the phase space $T^{*}Q$ has the coordinate chart $\{\theta_1,\theta_2,...,\theta_N,p_1,p_2,...,p_N\}$ and the Hamiltonian is given by
\begin{equation}\label{balance}
	\mathcal{H}=\frac{1}{2}\sum_{j=1}^{N}\frac{p_{j}^{2}}{I_{j}}+\Pi\left(\theta_2-\theta_1,\theta_3-\theta_2,...,\theta_{N}-\theta_{N-1}\right)\,.
\end{equation}
The dynamical equations follow from the Hamiltonian and read $\dot{\mathbf{X}}=\mathbf{J}\nabla_{\mathbf{X}}\mathcal{H}$, where 
\begin{equation}
	\mathbf{X}=\begin{bmatrix}
	\theta_{1} \\ \theta_{2} \\ \vdots \\ \theta_{N} \\ p_{1} \\ p_{2} \\ \vdots  \\ p_{N}
	\end{bmatrix}\,,
\end{equation}
and $\mathbf{J}$ is the following symplectic matrix
\begin{equation}
	\mathbf{J}=\begin{bmatrix}
	\mathbf{O}_{N} & \mathbf{I}_{N}\\
	-\mathbf{I}_{N} & \mathbf{O}_{N}
	\end{bmatrix}\,.
\end{equation}
$\mathbf{I}_N=[\delta_{ij}]$ is the $N\times N$ identity matrix, $\mathbf{O}_{N}$ is the $N\times N$ null matrix, and $\delta_{ij}$ is the Kronecker tensor. In particular,
\begin{equation}
	\dot{\theta}_{j}=\frac{p_{j}}{I_{j}}\,,\qquad\dot{p}_{j}=-\frac{\partial \Pi}{\partial\theta_{j}}\,,\qquad j=1,\hdots N\,,
\end{equation}
and from \eqref{angulmomemN} the conserved angular momentum is written as
\begin{equation}
    \mathsf{A}=\sum_{j=1}^{N}p_j(t)\,.
\end{equation}
The associated symplectic $1$ and $2$-forms are written as
\begin{equation} \label{forms}
	\alpha=\sum_{j=1}^{N}p_{j}\,d\theta_{j},\qquad d\alpha=\sum_{j=1}^{N}dp_{j}\wedge d\theta_{j}\,.
\end{equation}
The total kinetic energy of the elastic $N$-rotor is given by integrating the $1$-form $\alpha$:
\begin{equation}
    \mathsf{E}(t)=\int_0^t\,\alpha\,d\tau=\sum_{j=1}^{N}\int_0^t\,p_{j}(\tau)\,\dot{\theta}_j(\tau)\,d\tau=\sum_{j=1}^{N} \frac{1}{2} I_j \dot{\theta}^2_j(t)\,.
\end{equation}
To reveal the geometric nature of the dynamics, we consider the shape configuration space $Q_s$, which has the coordinate chart $\{\theta_1,\psi_2,\psi_3,...\psi_N\}$, where the shape parameters~$\psi_j=\theta_j-\theta_1$ represent the relative angular displacement of the $N-1$ rotors with respect to the first rotor. Since the total angular momentum $p_1+p_2+...+p_N=\mathsf{A}$ is known a priori, then $p_1=\mathsf{A}-p_2-p_3-\hdots-p_N$ and the motion must occur on the subspace $T^{*}Q/\mathsf{A}$, which has the coordinate chart $\{\theta_1,\psi_2,p_2,\psi_3,p_3,\hdots,\psi_N,p_N\}$, where $(p_j,\psi_j)$ are a pair of conjugate variables. The $1$-form in \eqref{forms} reduces to read
\begin{equation} \label{1formred2}
	\alpha=\mathsf{A} \,d\theta_1 + \sum_{j=2}^N p_j \,d\psi_j\,,
\end{equation}
and the associated symplectic $2$-form is written as
\begin{equation}
	d\alpha= \sum_{j=2}^N dp_j\wedge d\psi_j\,. 
\end{equation}
The reduced phase space $\mathcal{P}=T^*Q/\mathsf{A}$ has the geometric structure of a principal fiber bundle: the $2(N-1)$-dimensional shape manifold $\mathcal{B}$ with a coordinate chart $\{\psi_2,p_2,\cdots,\psi_N,p_N\}$ and transversal one-dimensional fibers $\mathcal{F}$ with coordinate chart $\{\theta_1\}$. 

The vector field $\dot{\mathbf{X}}_s=\dot{\mathbf{X}}_{\mathcal{B}}+\dot{\mathbf{X}}_{\mathcal{F}}$ can be decomposed as the sum of the flow 
\begin{equation}
	\dot{\mathbf{X}}_\mathcal{B}=\begin{bmatrix}
	   \dot{\psi}_2 \\ \dot{p}_2\\ \vdots  \\
    \dot{\psi}_{N} \\ \dot{p}_{N}\\
	\end{bmatrix}\,,
\end{equation}
on the shape manifold $\mathcal{B}$ and the flow $\dot{\mathbf{X}}_{\mathcal{F}}=\dot{\theta_1}$ along the fiber $\mathcal{F}$. Note that the motion on the shape manifold $\mathcal{B}$ is independent from that along the fiber. As a matter of fact, $\dot{\mathbf{X}}_{\mathcal{B}}$ does not depend on $\dot{\theta}_1$:
\begin{equation}
	\dot{\psi_{j}}=p_{j}\left(\frac{1}{I_{1}}+\frac{1}{I_{j}}\right)-\frac{1}{I_{1}}\left(\mathsf{A}-\sum_{k=2}^{N}p_{k}\right)\,,
	\qquad\dot{p}_{j}=\frac{\partial \hat{\Pi}}{\partial\psi_{j}}\,,
\end{equation}
and the associated reduced Hamiltonian is given by
\begin{equation}\label{Hamilreduced}
	\mathcal{H}_R(\psi_2,\cdots,\psi_N,p_2,\cdots,p_N)=\frac{1}{2 I_{1}}\left(\mathsf{A}-\sum_{k=2}^{N}p_{k}\right)^2+\sum_{j=2}^{N}\frac{p_{j}^{2}}{2 I_{j}}+\Pi\left(\psi_2,\psi_3,...,\psi_{N}\right)\,,
\end{equation}
where the potential is now given by
\begin{equation}
    \hat{\Pi}\left(\psi_2,\psi_3,\cdots,\psi_{N}\right)=\Pi_2(\psi_2)+\Pi_3(\psi_2-\psi_1)+\cdots+\Pi_{N}(\psi_{N}-\psi_{N-1})\,.
\end{equation}
On the contrary, the motion along the fiber depends on $\dot{X}_{\mathcal{B}}$ since
\begin{equation}
	\dot{\theta}_{1}=\frac{1}{I_{1}}\left(\mathsf{A}-\sum_{k=2}^{N}p_{k}\right)\,.
\end{equation} 
The motion in the reduced state space $\mathcal{P}=T^*Q/\mathsf{A}$ decouples in a reduced motion $\dot{\mathbf{X}}_{\mathcal{B}}$ on the shape manifold~$\mathcal{B}$ and a drift $\dot{\mathbf{X}}_\mathcal{F}$ along the fibers. The reduced motion on $\mathcal{B}$ is the shape-changing evolution of the connected rotors. Such a shape dynamics induces the rotors to rigidly rotate together by the varying angle $\theta_1$. From \eqref{1formred2}, we define the $1$-form $\widetilde{\alpha}=\alpha/\mathsf{A}$ and the total drift $\theta_1$ along the fiber follows by integrating the 1-form 
\begin{equation}
	d\theta_{1}=\widetilde{\alpha}-\sum_{k=2}^{N}\frac{p_{k}}{\mathsf{A}}\,d\psi_{k}\,,
\end{equation} 
that is 
\begin{equation}
	\theta_{1}=\int d\theta_{1}
	=\int_{0}^{t}\widetilde{\alpha}\,dt-\sum_{k=2}^{N}\int_{\gamma} \frac{p_{k}}{\mathsf{A}}\,d\psi_{k}\,,
\end{equation}
where $\gamma$ is a closed trajectory of the motion up to time $t$ in the shape manifold $\mathcal{B}$. Thus, 
\begin{equation}
	\theta_{1}(t)=\theta_{\text{dyn}}(t)+\theta_{\text{geom}}(t)\,,  
\end{equation}
where the dynamical and geometric rotation drifts are defined as
\begin{equation} \label{drift2}
	\theta_{\text{dyn}}(t)=\int_{0}^{t}\widetilde{\alpha}\,d\tau,\qquad
	\theta_{\text{geom}}(t)=-\sum_{k=2}^{N}\int_{\gamma} \frac{p_{k}}{\mathsf{A}}\,d\psi_{k}\,.
\end{equation}
Here, the dynamical rotation drift $\theta_{\text{dyn}}$ depends on the inertia of the rotors and can be written as
\begin{equation}
	\theta_{\text{dyn}}(t)=2\int_{0}^{t}\frac{\mathsf{K}(\tau)}{\mathsf{A}}\,d\tau\,,\qquad\mathsf{K}(t)=\frac{1}{2}\sum_{j=1}^{N} \frac{p_j^2}{I_j}\,,
\end{equation}
where $\mathsf{K}(t)$ is the total kinetic energy and $\mathsf{A}$ is the total angular momentum. If all the rotors are rigidly connected and cannot change their shape, i.e., $\dot{\psi_j}=0$ and so no motion on the base manifold $\mathcal{B}$, then the rotation drift is solely due to the inertia of the system measured by the total angular momentum. If the rotors  undergo changes in shape, i.e., the angles $\psi_j$ vary over time, then the motion on $\mathcal{B}$ induces the geometric rotation drift, which from \eqref{drift2} can be written as 
\begin{equation} \label{geomN}
	\theta_{\text{geom}}=-\sum_{k=2}^{N}\int_{\gamma}\frac{p_k}{\mathsf{A}} \,d\psi_k
	=-\sum_{k=2}^{N}\int_{S(\gamma)}\frac{1}{\mathsf{A}} \,dp_k\wedge\, d\psi_k\,.
\end{equation}
 Such a rotation drift is proportional to the area $S(\gamma)$ enclosed by the path $\gamma$ spanned by the motion on the shape manifold $\mathcal{B}$. Thus, it is purely geometric since it does not depend on the time it takes for the rotors to undergo a cyclic shape change, or to span the closed path $\gamma$ on $\mathcal{B}$. The part of kinetic energy that arises from the same cyclic shape change
\begin{equation} \label{geomN2}
\mathsf{E}_{\text{geom}}(t)=-\int_0^t\mathsf{A}\,\dot{\theta}_{\text{geom}}(\tau)\,d\tau=\sum_{k=2}^{N}\int_{\gamma}p_k\,d\psi_k
	=\sum_{k=2}^{N}\int_{S(\gamma)} dp_k\wedge\, d\psi_k\,,
\end{equation}
does not depend on the duration of the cyclic change either. Note that the energy difference 
\begin{equation}
    \mathsf{E}(t)-\mathsf{E}_{\text{geom}}(t)=\int_0^t\mathsf{A}\,\dot{\theta}_1\,d\tau\,,
\end{equation}
is that relative to the total rotation drift $\theta_1$. In the following, we will show that the base manifold can be endowed with a Riemannian structure. 

\subsection{Curvature and intrinsic metric of the shape manifold}\label{Sec:metricNpendulum}

The geometric rotation drift can be interpreted as the curvature of the shape manifold $\mathcal{B}$ equipped with a pseudo-Riemannian diagonal metric of the form
\begin{equation}
    ds^2=\sum_{j=2}^N \left[\epsilon_{p_j} G_{p_j} (d p_j)^2 + \epsilon_{\psi_j} G_{\psi_j} (d \psi_j)^2\right]\,,
\end{equation}
where the $2(N-1)$ non-negative metric coefficients (at least one being positive) depend on the coordinates $\{p_2,\psi_2,\cdots,p_N,\psi_N\}$, in general, and $(\epsilon_{p_2},\epsilon_{\psi_2},\cdots,\epsilon_{p_N},\epsilon_{\psi_N})$ is the signature of the manifold. The metric coefficients will be calculated using Cartan's structural equations as follows. From \eqref{geomN} the geometric drift follows by integrating the $2$-form
\begin{equation}\label{geom3}
d\,\widetilde{\alpha}=\sum_{j=2}^{N}d\,\widetilde{\alpha}_j\,,
\end{equation}
where 
\begin{equation}\label{connformalpha}
   \widetilde{\alpha}_j=-\frac{p_j}{\mathsf{A}} \,d\psi_j\,,\qquad 
   d\,\widetilde{\alpha}_j=-\frac{1}{\mathsf{A}} \,dp_j\wedge\, d\psi_j\\,\qquad j=2,\cdots,N\,.
\end{equation}
Drawing on Cartan's second structural equations~\eqref{SecondCartan}, the collection of the $(N-1)$ $2$-forms $d\,\widetilde{\alpha}_j$ are interpreted as the non-zero curvature $2$-forms of a $2(N-1)$-dimensional manifold, and the associated connection $1$-forms are $\widetilde{\alpha}_j$. For a metric-compatible connection on the $M=2(N-1)$-dimensional shape manifold there are $M(M-1)/2=(N-1)(2 N -3)$ connection $1$-forms and as many curvature $2$-forms. In particular, 
\begin{equation}\label{zerocon}
     \omega^{p_j}{}_{p_k}\,,~ 
\omega^{\psi_j}{}_{\psi_k}\,,\qquad j<k=2,\hdots,N\,,
\end{equation}
are $2\times \frac{(N-1)(N-2)}{2}=(N-1)(N-2)$ connection $1$-forms. The remaining $(N-1)^2$ connection $1$-forms are
\begin{equation}\label{nonzerocon}
    \omega^{\psi_k}{}_{p_j}\,,\qquad j,\,k=2,\hdots, N\,.
\end{equation}
Therefore, in total we have $(N-1)(N-2)+(N-1)^2=(N-1)(2N-3)$ connection $1$-forms and as many curvature $2$-forms given by
\begin{equation}
\begin{aligned}
     & \mathcal{R}^{p_j}{}_{p_k}= 
\mathcal{R}^{\psi_j}{}_{\psi_k}=0\,,&& j<k=2,\hdots,N\,,\\
    & \mathcal{R}^{\psi_k}{}_{p_j}=0\,,&&j\neq k\,,\\
    & \mathcal{R}^{\psi_j}{}_{p_j}=d\,\widetilde{\alpha}_j=-\frac{1}{\mathsf{A}} \,dp_j\wedge\, d\psi_j\,,&&j=2,\hdots, N\,.
\end{aligned}
\end{equation}
The unknown connection $1$-forms satisfy Cartan's second structural equations~\eqref{SecondCartan}
\begin{equation}
\begin{aligned}
     & \mathcal{R}^{p_j}{}_{p_k}=0=d\omega^{p_j}{}_{p_k}+\omega^{p_j}{}_{\gamma}\wedge\omega^{\gamma}{}_{p_k} \,,&& j<k=2,\hdots,N\,,\\
     & \mathcal{R}^{\psi_j}{}_{\psi_k}=0=d\omega^{\psi_j}{}_{\psi_k}+\omega^{\psi_j}{}_{\gamma}\wedge\omega^{\gamma}{}_{\psi_k}\,,&& j<k=2,\hdots,N\,,\\
    & \mathcal{R}^{\psi_k}{}_{p_j}=0=d\omega^{\psi_k}{}_{p_j}+\omega^{\psi_k}{}_{\gamma}\wedge\omega^{\gamma}{}_{p_j}\,,&&j\neq k\,,\\
    & \mathcal{R}^{\psi_j}{}_{p_j}=-\frac{1}{\mathsf{A}} \,dp_j\wedge\, d\psi_j=d\omega^{\psi_j}{}_{p_j}+\omega^{\psi_j}{}_{\gamma}\wedge\omega^{\gamma}{}_{p_j}\,,&&j=2,\hdots, N\,.
\end{aligned}
\end{equation}
These are more explicitly written  as
\begin{equation}
\begin{aligned}
     & d\omega^{p_j}{}_{p_k}+\sum_{i=2}^N\omega^{p_j}{}_{\psi_i}\wedge\omega^{\psi_i}{}_{p_k}+\sum_{i=2}^N\omega^{p_j}{}_{p_i}\wedge\omega^{p_i}{}_{p_k}=0 \,,&& j<k=2,\hdots,N\,,\\
     & d\omega^{\psi_j}{}_{\psi_k}+\sum_{i=2}^N\omega^{\psi_j}{}_{\psi_i}\wedge\omega^{\psi_i}{}_{\psi_k}+\sum_{i=2}^N\omega^{\psi_j}{}_{p_i}\wedge\omega^{p_i}{}_{\psi_k}=0\,,&& j<k=2,\hdots,N\,,\\
    & d\omega^{\psi_k}{}_{p_j}+\sum_{i=2}^N\omega^{\psi_k}{}_{\psi_i}\wedge\omega^{\psi_i}{}_{p_j}+\sum_{i=2}^N\omega^{\psi_k}{}_{p_i}\wedge\omega^{p_i}{}_{p_j}=0\,,&&j\neq k\,,\\
    & d\omega^{\psi_j}{}_{p_j}+\sum_{i=2}^N\omega^{\psi_j}{}_{\psi_i}\wedge\omega^{\psi_i}{}_{p_j}+\sum_{i=2}^N\omega^{\psi_j}{}_{p_i}\wedge\omega^{p_i}{}_{p_j}=-\frac{1}{\mathsf{A}} \,dp_j\wedge\, d\psi_j\,,&&j=2,\hdots, N\,.
\end{aligned}
\end{equation}

The case $N=2$ is trivial as there is a unique solution $\omega^{\psi_2}{}_{p_2}=\frac{\psi_2}{\mathsf{A}}\,dp_2+\xi$ given in \eqref{1form3}, where $\xi$ is any closed $1$-form, which can be neglected as it does not contribute to the geometric phase. For $N>2$ we have a system of nonlinear equations to solve for the connection $1$-forms, and there may be more than one solution. If we require the only non-zero connection forms to be $\widetilde{\alpha}_j,\,\,j=2,\hdots,N$ in \eqref{connformalpha} then we have a solution\footnote{One can add arbitrary closed $1$-forms to each connection $1$-form, but these do not yield new solutions since the difference is a closed $1$-form. They can be neglected as they are not physically relevant. As a matter of fact, they do not contribute to the geometric phase as their integrals over any closed curve vanish. Thus, the freedom to add an arbitrary closed $1$-form is physically inconsequential and it does not give any new solutions.} that follows from \eqref{zerocon} and~\eqref{nonzerocon} as $\omega^{\psi_j}{}_{p_j}=-\frac{p_j}{\mathsf{A}} \, d\psi_j$, $j=2,\hdots,N$, and $\omega^{p_k}{}_{\psi_j}=\omega^{p_k}{}_{p_j}=\omega^{\psi_k}{}_{\psi_j}=0$. From \eqref{geom3}, it follows that the non-zero curvature 2-forms are the exterior derivatives of the $1$-forms $\widetilde{\alpha}_j$
\begin{equation}\label{Rpsij_pj}
    \mathcal{R}^{\psi_j}{}_{p_j}=d\,\omega^{\psi_j}{}_{p_j}=d\,\widetilde{\alpha}_j\,,\qquad j=2,\hdots, N\,.
\end{equation}
From \eqref{Levi-Civita-connection-forms} we then have
\begin{equation} 
\begin{aligned}
	& \omega^{p_k}{}_{\psi_j}=\frac{\partial_{\psi_j}\mathsf{G}_{p_k}}{2\sqrt{\mathsf{G}_{p_k}\mathsf{G}_{\psi_j}}}\,dp_k
	-\epsilon_{p_k}\epsilon_{\psi_j}\,\frac{\partial_{p_k}\mathsf{G}_{\psi_j}}{2\sqrt{\mathsf{G}_{p_k}\mathsf{G}_{\psi_j}}}\,d\psi_j
	=0\,,&& j\neq k\,, \\
    & \omega^{\psi_k}{}_{\psi_j}=\frac{\partial_{\psi_j}\mathsf{G}_{\psi_k}}{2\sqrt{\mathsf{G}_{\psi_k}\mathsf{G}_{\psi_j}}}\,d\psi_k
	-\epsilon_{\psi_k}\epsilon_{\psi_j}\,\frac{\partial_{\psi_k}\mathsf{G}_{\psi_j}}{2\sqrt{\mathsf{G}_{\psi_k}\mathsf{G}_{\psi_j}}}\,d\psi_j
	=0\,,&& j<k=2,\hdots,N\,, \\
    & \omega^{p_k}{}_{p_j}=\frac{\partial_{p_j}\mathsf{G}_{p_k}}{2\sqrt{\mathsf{G}_{p_k}\mathsf{G}_{p_j}}}\,dp_k
	-\epsilon_{p_k}\epsilon_{p_j}\,\frac{\partial_{p_k}\mathsf{G}_{p_j}}{2\sqrt{\mathsf{G}_{p_k}\mathsf{G}_{p_j}}}\,d p_j
	=0\,,&& j<k=2,\hdots,N\,,\\
  & \omega^{p_j}{}_{\psi_j}=\frac{\partial_{\psi_j}\mathsf{G}_{p_j}}{2\sqrt{\mathsf{G}_{p_j}\mathsf{G}_{\psi_j}}}\,dp_j
	-\epsilon_{p_j}\epsilon_{\psi_j}\,\frac{\partial_{p_j}\mathsf{G}_{\psi_j}}{2\sqrt{\mathsf{G}_{p_j}\mathsf{G}_{\psi_j}}}\,d\psi_j
	=-\frac{p_j}{\mathsf{A}} \,d\psi_j\,,&& j=2,\hdots,N\,.\\
\end{aligned}
\end{equation}
Thus, we must have 
\begin{equation}\label{G1G2eqN}
	\partial_{\psi_k}\mathsf{G}_{p_j}=0,\qquad \partial_{p_k}\mathsf{G}_{p_j}=0,\qquad
	\partial_{\psi_k}\mathsf{G}_{\psi_j}=0,\qquad \partial_{p_k}\mathsf{G}_{\psi_j}=0,\qquad k\neq j\,,
\end{equation}
and
\begin{equation}\label{G1G2eqN}
	\partial_{\psi_j}\mathsf{G}_{p_j}=0,\qquad -\epsilon_{p_j}\epsilon_{\psi_j}\,\frac{\partial_{p_j}\mathsf{G}_{\psi_j}}{2\sqrt{\mathsf{G}_{p_j}\mathsf{G}_{\psi_j}}}=-\frac{p_j}{\mathsf{A}}\,.
\end{equation}
The above relations imply that $\mathsf{G}_{p_j}=\mathsf{G}_{p_j}(p_j)$ and $\mathsf{G}_{\psi_j}=\mathsf{G}_{\psi_j}(\psi_j,p_j)$. Thus, the metric coefficients $G_{p_j}$ and $G_{\psi_j}$ depend only on the coordinates $\{\psi_j,p_j\}$ of the submanifold (hyper-plane) $\mathcal{B}_j$. The curvature $2$-forms $\mathcal{R}^{\psi_j}{}_{p_j}$ in \eqref{Rpsij_pj} are now expressed in terms of the metric coefficients using \eqref{R12curvatureform} as
\begin{equation}
   \mathcal{R}^{\psi_j}{}_{p_j}=d\omega^{\psi_j}{}_{p_j} = K_{(p_j,\psi_j)}\sqrt{\mathsf{G}_{p_j}\mathsf{G}_{\psi_j}}\,dp_j\wedge d\psi_j\,,
\end{equation}
where $K_{(p_j,\psi_j)}$ is the Gaussian curvature of the hyper-plane $\mathcal{B}_j$ with coordinates $\{\psi_j,p_j\}$~(see \eqref{Gaussiancurvature}). Then, \eqref{Rpsij_pj} imposes the equality of the $2$-forms $d\,\widetilde{\alpha}_j=d\omega^{\psi_j}{}_{p_j}$, that is
\begin{equation}\label{QE}
    -\frac{1}{\mathsf{A}} \,dp_j\wedge d\psi_j = K_{(p_j,\psi_j)}\sqrt{\mathsf{G}_{p_j}\mathsf{G}_{\psi_j}}\,dp_j\wedge d\psi_j\,,
\end{equation}
and it follows that
\begin{equation}\label{QE}
   K_{(p_j,\psi_j)}= -\frac{\sqrt{\mathsf{G}_{p_j}\mathsf{G}_{\psi_j}}}{\mathsf{A}}\,.
\end{equation}
Similar to the elastic double rotor, we further require that the symplectic $2$-form $d\,\widetilde{\alpha}_j$ be compatible with the (pseudo) Riemannian volume~(area) $2$-form $\mathsf{vol}_{\mathcal{B}_j}=\sqrt{\mathsf{G}_{p_j}\mathsf{G}_{\psi_j}} \,dp_j\wedge d\psi_j$ of the submanifold $\mathcal{B}_j$, that is
\begin{equation}
    \frac{1}{|\mathsf{A}|} \,dp_j\wedge d\psi_j = \sqrt{\mathsf{G}_{p_j}\mathsf{G}_{\psi_j}}\,dp_j\wedge d\psi_j\,.
\end{equation}
This implies that
\begin{equation}\label{canc}
    \frac{1}{|\mathsf{A}|}=\sqrt{\mathsf{G}_{p_j}\mathsf{G}_{\psi_j}}\,,
\end{equation}
which together with \eqref{G1G2eqN}$_2$ gives us
\begin{equation}
    \epsilon_{p_j}\epsilon_{\psi_j}\,\partial_{p_j}\mathsf{G}_{\psi_j}=2\frac{p_j}{\mathsf{A}|\mathsf{A}|}\,.
\end{equation}
Solving for $\mathsf{G}_{\psi_j}$, and using \eqref{canc} to solve for $\mathsf{G}_{p_j}$ we get
\begin{equation}\label{coeffN}
	\mathsf{G}_{p_j}=\frac{1}{{p_j^2+\mu_j^2}},\qquad \mathsf{G}_{\psi_j}=\frac{p_j^2+\mu_j^2}{\mathsf{A}^2},\qquad \epsilon_{p_j}=\mathrm{sgn}(\mathsf{A}),\qquad \epsilon_{\psi_j}=1\,, 
\end{equation}
where we have imposed that both metric coefficients are positive and $\mu_j$ are arbitrary constants. The shape manifold $\mathcal{B}$ is thus the product manifold of $(N-1)$ shape submanifolds $\mathcal{B}_j$ with local coordinate charts $\{p_j,\psi_j\}$, or $\mathcal{B}=\mathcal{B}_2\times \cdots\times\mathcal{B}_N$ (see \S\ref{Sec:ProductSpace}).

Each submanifold is the shape space of two adjacent rotors, or double rotor. Thus, the intrinsic metric of each submanifold follows from \eqref{coeffN}, (or from \eqref{metric2}) as
\begin{equation}\label{metric3}
	\mathbf{G}_{j}=\frac{\mathrm{sgn}(\mathsf{A})}{p_j^2+\mu_j^2}\,dp_j\otimes d p_j+\frac{p_j^2+\mu_j^2}{\mathsf{A}^2}\,d\psi_j\otimes d\psi_j\,.
\end{equation}
Then the metric of $\mathcal{B}$ is the product of these metrics, i.e.,
\begin{equation}\label{metricN}
  \mathbf{G}=\mathbf{G}_2\times\hdots\times\mathbf{G}_N=\sum_{j=2}^N \left[\frac{\mathrm{sgn}(\mathsf{A})}{p_j^2+\mu_j^2}\,dp_j\otimes d p_j+\frac{p_j^2+\mu_j^2}{\mathsf{A}^2}\,d\psi_j\otimes d\psi_j\right]\,. 
\end{equation}
From \eqref{geom3} the geometric drift follows by integrating the $2$-form
\begin{equation}
d\,\widetilde{\alpha}= \sum_{j=2}^N\,\mathcal{R}^{p_j}{}_{\psi_j}(\mathbf{e}_{p_j},\mathbf{e}_{\psi_j})\,.
\end{equation}
This is the sum of the curvature $2$-forms of each submanifold $\mathcal{B}_j$, that is 
\begin{equation} \label{geom4}
	\theta_{\text{geom}}=-\sum_{j=2}^{N}\int_{S(\gamma)} \mathcal{R}^{p_j}{}_{\psi_j}(\mathbf{e}_{p_j},\mathbf{e}_{\psi_j})=-\sum_{j=2}^{N}\int_{S(\gamma)}\frac{1}{\mathsf{A}} \,dp_j\wedge\, d\psi_j\,,
\end{equation}
where each term is both the oriented area and curvature of the projected path $\gamma$ on the hyper-plane $\mathcal{B}_j$ with coordinates $\{\psi_j,p_j\}$. The geodesics of the product manifold $\mathcal{B}$ are the Cartesian product of the geodesics of each submanifold $\mathcal{B}_j$, which follow from \eqref{geodesic}. 

\textcolor{black}{Without lose of generality, one has the freedom to define the sign of the total angular momentum as either positive or negative, e.g., counterclockwise or clockwise, and viceversa. The base manifold~$\mathcal{B}$ can then be endowed with two distinct metrics both compatible with the geometric phase. In the following, we will show that $\mathcal{B}$ is the product manifold of $N-1$ hyperbolic planes $\mathbb{H}^2$~($\mathsf{A}>0$), or Robertson-Walker $2$D spacetimes~($\mathsf{A}<0$) depending on the convection used to define the rotation sign of the total angular momentum~$\mathsf{A}$.} 

\begin{remark}[Metric Uniqueness]\label{Nrotoruniquemetric}
\textcolor{black}{Similarly to the double rotor problem~(see Remark~\ref{2rotoruniquemetric}), a unique metric  can be defined by matching the symplectic forms~$\beta_j=-\mathsf{A}\,\widetilde{\alpha}_j=p_j\, d\psi_j$ of the reduced dynamics on~$\mathcal{B}$~(see \eqref{connformalpha}) with the connection $1$-forms $\widetilde{\alpha}_j$ in \eqref{zerocon}. As a result, the geometric phase is directly linked to curvature, and the constant of proportionality in this relationship is given by $-\frac{1}{\mathsf{A}}$ as indicated by \eqref{geomN}. Such a matching equips $\mathcal{B}$ with the following pseudo-Riemannian metric
\begin{equation}\label{metricN2}
  \mathbf{G}=\mathbf{G}_2\times\hdots\times\mathbf{G}_N=\sum_{j=2}^N \left[-\frac{1}{p_j^2+\mu_j^2}\,dp_j\otimes d p_j+(p_j^2+\mu_j^2)\,d\psi_j\otimes d\psi_j\right]\,, 
\end{equation}
which is disguised metric of a multi-universe of $2$D Robertson-Walker spacetimes, $\forall \mu_j\in\mathbb{R}$, $j=2,\hdots,N$ as shown in the following.}
\end{remark}

\subsection{Negative angular momentum: Multi-universe}

Choosing $\mathsf{A}<0$, the metric~\eqref{metricN} describes a multi-universe  of expanding $2$D Robertson-Walker spacetime universes~\citep{GRAVITATION,Carroll2003spacetime}. Indeed, for $\mu_j\neq 0$ we use the chart coordinate transformation~\eqref{coordchange1}
\begin{equation}\label{coordchange4}
   t_j=\tanh^{-1}\left[\frac{p_j}{\sqrt{p_j^2+\mu_j^2}}\right],\qquad x_j=\psi_j\,,\qquad j=2,\cdots, N\,. 
\end{equation}
Then $dt_j=d p_j/\sqrt{p_j^2+\mu_j^2}$, ~$dx_j=d\psi_j$, and the metric~\eqref{metricN} transforms into the sum of $(N-2)$ $2$-dimensional Robertson-Walker metrics of each submanifold $\mathcal{B}_j$
\begin{equation}
    ds^2=\sum_{j=2}^N ds^2_j,\quad\quad\,ds^2_j=-dt_j^2 + \frac{\mu_j^2 (\cosh{t_j})^2}{\mathsf{A}^2} \,dx_j^2\,,
    \qquad j=2,\cdots, N\,,
\end{equation}
with the scale factor $a(t_j)\sim\cosh{t_j}$. The associated Hubble constants of each spacetime is $H_j=\dot{a}_j/a_j=\tanh(t_j)$, indicating a matter-dominated universe for small $t_j$ and vacuum-dominated for large $t_j$~\citep{Carroll2003spacetime}. Similarly, if $\mu_j=0$ we use the coordinate transformation~\eqref{coordchange2}:
\begin{equation}\label{coordchange5}
   t_j=\mathrm{e}^{p_j},\qquad x_j=\psi_j\,,\qquad j=2,\cdots, N\,, 
\end{equation}
where $dt_j=d p_j/p_j$ and $ dx_j=d\psi_j$, and the metric transforms to
\begin{equation}
    ds^2_j=-dt_j^2 + \frac{\mathrm{e}^{2t_j}}{\mathsf{A}^2} dx_j^2\,,\qquad j=2,\cdots, N\,,
\end{equation}
which is still a Robertson-Walker metric with the scale factor $a(t_j)\sim\mathrm{e}^{t_j}$ and Hubble constant $H=\dot{a}_j/a_j=1$~\citep{Carroll2003spacetime}.

\subsection{Positive angular momentum: The hyperbolic product space $\mathbb{H}^{2(N-1)}$}

\textcolor{blue}{Choosing} $\mathsf{A}>0$, each submanifold $\mathcal{B}_j$ is a hyperbolic plane $\mathbb{H}^2$ and the shape manifold is the Cartesian product of $N-1$ hyperbolic planes $\mathbb{H}^2$, each with negative Gaussian curvature $\mathsf{K}_j=-1$. So, $\mathcal{B}$ is the hyperbolic product space $\mathbb{H}^{2(N-1)}$. As an example, for $\mu=0$ we use the coordinate transformations~\eqref{coordchange1},~\eqref{coordchange2} and the metric~\eqref{metricN}  transforms into the sum of $N-2$ metrics
\begin{equation} \label{Surface-Revolution2}
   ds^2=\sum_{j=2}^N ds^2_j,\quad\quad\,ds^2_j=dt_j^2 + R_j^2(t) \,dx_j^2\,,\qquad j=2,\cdots, N\,,
\end{equation}
where $R_j(t)=\frac{1}{\mathsf{A}^2}\mathrm{e}^{2\,t_j}$. The metrics $ds^2_j$ of the submanifolds $\mathcal{B}_j$ are disguised metrics of the hyperbolic plane as the change of coordinates $\tilde{x}_j=x_j/\sqrt{\mathsf{A}}$ and $\tilde{y}_j=\mathsf{A}\exp(-t_j)$ transform each of them into
\begin{equation} \label{hypermetricN}
    ds^2_j=\frac{d\tilde{x}_j^2+d\tilde{y}_j^2}{\tilde{y}_j^2}\,,\qquad j=2,\cdots, N\,.
\end{equation}

\section{Dynamics of nonlinear elastic $N$-rotors under self-equilibrated external moments}

We next generalize the elastic $N$-rotor system described above by assuming that time-dependent external moments $\mathsf{M}^e_j(t)$, $j=1,\cdots,N$ act on the rigid rotors (see Fig.~\ref{N-pendulum}). In order to preserve the invariance of the total angular momentum we assume that
\begin{equation}\label{balance2}
    \sum_{j=1}^N\mathsf{M}^e_j(t)=0\,. 
\end{equation}
The associated Lagrangian is\footnote{The external moments appear in the Lagrange d'Alembert principle. Equivalently, one can use Hamilton's principle using the modified Lagrangian given in \eqref{LagrN2}.}
\begin{equation} \label{LagrN2}
    \mathcal{L}=\sum_{j=1}^N\frac{1}{2}I_{j}\,\dot{\theta}_{j}^{2}- \Pi\left(\theta_2-\theta_1,\theta_3-\theta_2,...,\theta_{N}-\theta_{N-1}\right) +\sum_{j=1}^N\mathsf{M}^e_j(t)\,\theta_j\,,
\end{equation}
and the associated dynamical equations follow by extremizing the action $\int \mathcal{L} dt $ as
\begin{equation} \label{LagrangeeqsN1}
 	\frac{d}{dt} \frac{\partial\mathcal{L}}{\partial\dot{\theta}_j}
	-\frac{\partial\mathcal{L}}{\partial\theta_{j}}=I_j\ddot{\theta}_j+\frac{\partial \Pi}{\partial\theta_j}-\mathsf{M}^e_j(t)=0\,,
	\quad j=1,\cdots N\,.  
\end{equation}
From~\eqref{balanceN} potential moments are in equilibrium and summing up equations~\eqref{LagrangeeqsN1} yields 
\begin{equation} 
	\frac{d}{dt} \sum_{j=1}^N I_j\dot{\theta}_j(t)=\sum_{j=1}^N\mathsf{M}^e_j(t)\,.
\end{equation}
From \eqref{balance2} the sum of the external moments on the right-hand side is null and the total angular momentum is conserved, i.e.,
\begin{equation}\label{angulmomemN}
    I_{1}\dot{\theta}_{1}(t)+I_2\dot{\theta}_2(t)+\cdots+I_N\dot{\theta_N}(t)=\mathsf{A}\,.
\end{equation}

\subsection{Extended autonomous 
Hamiltonian system}

The elastic $N$-rotor is a non-autonomous system since the Lagrangian is explicitly time-dependent. We  can associate an extended autonomous Hamiltonian system on the cotangent bundle $T^*Q_t$ of the extended configuration space $Q_t=\mathbb{R}\times \mathbb{T}^N$, i.e., the Cartesian product of the real line $\mathbb{R}$ and the $N$-torus. $Q_t$ has the coordinate chart $\{t,\theta_1,\theta_2,\cdots,\theta_N\}$. The conjugate momentum of time $t$ is the energy $E$ and $p_j=I_j\dot{\theta}_j$ are the conjugate momenta of the angles $\theta_j$. Thus, the phase space $T^{*}Q_t$ is the cotangent space of $Q_t$, and has the coordinate chart $\{t,\theta_1,\theta_2,...,\theta_N,E,p_1,p_2,...,p_N\}$. A generic trajectory in the extended phase space is parameterized by the parameter $\lambda$. The Hamiltonian is given by
\begin{equation}\label{balance}
	\mathcal{H}=E+\frac{1}{2}\sum_{j=1}^{N}\frac{p_{j}^{2}}{I_{j}}+\Pi\left(\theta_2-\theta_1,\theta_3-\theta_2,...,\theta_{N}-\theta_{N-1}\right)-\sum_{j=1}^N\mathsf{M}^e_j\,\theta_j\,.
\end{equation}
The dynamical equations follow from the Hamiltonian by $\mathbf{X}^{\prime}=\mathbf{J}\nabla_{\mathbf{X}}\mathcal{H}$, where $\mathbf{X}^{\prime}=\frac{d \mathbf{X}}{d\lambda}$ denotes differentiation with respect to $\lambda$, and 
\begin{equation}
    \mathbf{X}=\begin{bmatrix}
        t \\ \theta_{1} \\ \theta_{2} \\ \vdots \\\theta_{N} \\ E \\ p_{1} \\ p_{2} \\ \vdots \\p_N
    \end{bmatrix}
    \,,
\end{equation}
and $\mathbf{J}$ is the symplectic matrix
\begin{equation}
	\mathbf{J}=\begin{bmatrix}
	\mathbf{O}_{N+1} & \mathbf{I}_{N+1}\\
	-\mathbf{I}_{N+1} & \mathbf{O}_{N+1}
	\end{bmatrix}\,.
\end{equation}
$\mathbf{I}_{N+1}=[\delta_{ij}]$ is the $N+1\times N+1$ identity matrix, $\mathbf{O}_{N+1}$ is the $N+1\times N+1$ null matrix, and $\delta_{ij}$ is the Kronecker tensor. In particular,
\begin{equation}
	t^{\prime}=\frac{\partial\mathcal{H}}{\partial E}=1\,,\quad E^{\prime}=-\frac{\partial\mathcal{H}}{\partial t}=\sum_{j=1}^N\frac{d\mathsf{M}^e_j}{d t}\theta_j\,,\quad\theta_j^{\prime}=\frac{p_{j}}{I_{j}}\,,\quad p_j^{\prime}=-\frac{\partial \Pi}{\partial\theta_{j}}+\mathsf{M}^e_j(t)\,,\quad j=1,\hdots N\,.
\end{equation}
From \eqref{angulmomemN} the conserved total angular momentum is $\mathsf{A}=\sum_{j=1}^{N}p_j$. The associated symplectic $1$ and $2$-forms are
\begin{equation} \label{formst}
	\alpha=E dt+\sum_{j=1}^{N}p_{j}\,d\theta_{j},\qquad d\alpha=dE\wedge dt + \sum_{j=1}^{N}dp_{j}\wedge d\theta_{j}\,.
\end{equation}

We now consider the shape configuration space $Q_s$, which has the coordinate chart $\{t,\theta_1,\psi_2,\psi_3,...\psi_N\}$, where the shape parameters~$\psi_j=\theta_j-\theta_1$ represent the relative angular displacement of the $N-1$ rigid rotors with respect to the first bar. Since the total angular momentum $p_1+p_2+...+p_N=\mathsf{A}$ is known a priori, then $p_1=\mathsf{A}-p_2-p_3-\hdots-p_N$ and the motion must occur on the subspace $T^{*}Q_t/\mathsf{A}$, which has the coordinate chart $\{t,E,\theta_1,\psi_2,p_2,\psi_3,p_3,\hdots,\psi_N,p_N\}$, where $(t,E)$  and $(p_j,\psi_j)$ are pairs of conjugate variables. The $1$-form in \eqref{forms} reduces to
\begin{equation} \label{1formred2a}
	\alpha= \mathsf{A} \,d\theta_1  + E dt+ \sum_{j=2}^N p_j \,d\psi_j\,,
\end{equation}
and the associated symplectic $2$-form is written as
\begin{equation}
	d\alpha=dE \wedge dt + \sum_{j=2}^N dp_j\wedge d\psi_j\,. 
\end{equation}
The reduced phase space $\mathcal{P}=T^*Q/\mathsf{A}$ has the geometric structure of a principal fiber bundle: the $2 N$-dimensional shape manifold $\mathcal{B}$ with coordinate chart $\{t,E,\psi_2,p_2,\cdots,\psi_N,p_N\}$ and transversal one-dimensional fibers $\mathcal{F}$ with coordinate $\{\theta_1\}$. 
The Hamiltonian vector field $\mathbf{X}_s^{\prime}$ in $\mathcal{P}=T^*Q/\mathsf{A}$ can be decomposed as the sum of the flow 
\begin{equation}
    \mathbf{X}^{\prime}_\mathcal{B}=\begin{bmatrix}
        t^{\prime}\\ E^{\prime} \\ \psi_2^{\prime} \\ p_2^{\prime}\\ \vdots \\ \psi_N^{\prime} \\ p_N^{\prime}
    \end{bmatrix}
    \,,
\end{equation}
on the shape manifold $\mathcal{B}$ and the flow $\mathbf{X}^{\prime}_{\mathcal{F}}=\theta_1^{\prime}$ along the fiber $\mathcal{F}$. 
The dynamics on the shape manifold $\mathcal{B}$ is governed by the reduced (time-varying) Hamiltonian 
\begin{equation}\label{Hamilreduced2}
	\mathcal{H}_R(t,E,\psi_2,p_2,\cdots,\psi_N,p_N)=E+\frac{1}{2 I_{1}}\left[\mathsf{A}-\sum_{k=2}^{N}p_{k}\right]^2+\sum_{j=2}^{N}\left[\frac{p_{j}^{2}}{2 I_{j}}-\mathsf{M}^e_j(t)\psi_j\right]+\Pi\left(\psi_2,\psi_3,...,\psi_{N}\right)\,,
\end{equation}
and the components of the Hamiltonian vector field $\mathbf{X}^{\prime}_{\mathcal{B}}$ are
\begin{equation}
	t^{\prime}=1\,,\quad E^{\prime}=-\frac{\partial\mathcal{H}_R}{\partial t}=\sum_{j=1}^N\frac{d\mathsf{M}^e_j}{d t}\psi_j\,,\quad \psi_j^{\prime}=p_{j}\left(\frac{1}{I_{1}}+\frac{1}{I_{j}}\right)-\frac{1}{I_{1}}\left(\mathsf{A}-\sum_{k=2}^{N}p_{k}\right)\,,
	\quad p_j^{\prime}=\frac{\partial \hat{\Pi}}{\partial\psi_{j}}-\mathsf{M}^e_j(t)\,.
\end{equation}
Notice that the motion along the fiber depends on $\mathbf{X}^{\prime}_{\mathcal{B}}$ since
\begin{equation}
	\theta_1^{\prime}=\frac{1}{I_{1}}\left(\mathsf{A}-\sum_{k=2}^N p_k\right)\,.
\end{equation} 
From \eqref{1formred2}, we define the $1$-form $\widetilde{\alpha}=\alpha/\mathsf{A}$ and the total drift $\theta_1$ along the fiber follows by integrating the form 
\begin{equation}
	d\theta_{1}=\widetilde{\alpha}-\frac{E}{\mathsf{A}} dt - \sum_{k=2}^{N}\frac{p_{k}}{\mathsf{A}} \,d\psi_{k}\,,
\end{equation} 
that is 
\begin{equation}
	\theta_{1}=\int d\theta_{1}
	=\int_{0}^{\lambda}\widetilde{\alpha}\,d\widetilde{\lambda}-\int_{\gamma} \left(\frac{E}{\mathsf{A}} dt + \sum_{k=2}^{N}\frac{p_{k}}{\mathsf{A}}\,d\psi_{k}\right)\,,
\end{equation}
where $\gamma$ is a closed trajectory of the motion  on the shape manifold $\mathcal{B}$ parameterized by $\lambda$. Thus, 
\begin{equation}
	\theta_{1}=\theta_{\text{dyn}}+\theta_{\text{geom}}\,,  
\end{equation}
where the dynamical and geometric rotation drifts are defined as
\begin{equation} \label{drift2a}
	\theta_{\text{dyn}}(\lambda)=\int_{0}^{\lambda}\widetilde{\alpha}\,d\widetilde{\lambda},\qquad
	\theta_{\text{geom}}(\lambda)=-\int_{\gamma} \left(\frac{E}{\mathsf{A}} dt + \sum_{k=2}^{N}\frac{p_{k}}{\mathsf{A}}\,d\psi_{k}\right)\,.
\end{equation}
Here, the dynamical rotation drift $\theta_{\text{dyn}}$ depends on the inertia of the elastic $N$-rotor and can be written as
\begin{equation}
	\theta_{\text{dyn}}(\lambda)=2\int_{0}^{\lambda}\frac{\mathsf{K}(\widetilde{\lambda})+E(\widetilde{\lambda})}{\mathsf{A}}\,d\widetilde{\lambda}\,,\qquad \mathsf{K}=\frac{1}{2}\sum_{j=1}^{N} \frac{p_j^2}{I_j}\,,
\end{equation}
where $\mathsf{K}$ and $\mathsf{A}$ are the total kinetic energy and the total angular momentum, respectively. If the rotors of the elastic $N$-rotor are rigidly connected and cannot change their shape, i.e., no motion on the shape manifold as  $\psi_j^{\prime}=0$, then the rotation drift is solely due to the inertia of the system measured by the total angular momentum and it is measured by $\theta_{\text{dyn}}$. If the elastic $N$-rotor changes its shape, i.e., the angles $\psi_j$ vary over time, then the motion on $\mathcal{B}$ induces also the geometric rotation drift $\theta_{\text{geom}}$. From \eqref{drift2a}, and using Stokes' theorem
\begin{equation} \label{geomN3}
	\theta_{\text{geom}}
	=-\int_{S(\gamma)}\left( \frac{1}{\mathsf{A}} dE\wedge\, dt + \sum_{k=2}^{N}\frac{1}{\mathsf{A}} dp_k\wedge\, d\psi_k \right)\,.
\end{equation}
 The geometric drift is thus proportional to the area $S(\gamma)$ enclosed by the path $\gamma$ spanned by the motion on the shape manifold $\mathcal{B}$. The $2$-form $dE \wedge dt$ encodes the effects of the time-dependent external moments on the geometric drift. The remaining $2$-forms are the same as those of a free elastic $N$-rotor given in \eqref{geom4} and measure the effects of the $N$-rotor shape changes. 
In the following, we will show that the base manifold~$\mathcal{B}$ can be endowed with a Riemannian structure.

\subsection{Curvature and intrinsic metric of the shape manifold}

One can interpret the geometric rotation drift in \eqref{geomN3} as the curvature of the $2 N$-dimensional shape manifold $\mathcal{B}$ equipped with a pseudo-Riemannian metric of the following form
\begin{equation}
    ds^2=\epsilon_t\, G_t \,dt^2 + \epsilon_E\, G_E \,dE^2 + \sum_{j=2}^N \left[(\epsilon_{p_j} G_{p_j} (d p_j)^2 + \epsilon_{\psi_j}G_{\psi_j} (d \psi_j)^2\right]\,,
\end{equation}
where the $2N$ non-negative metric coefficients (at least one being positive) depend on the coordinates $\{t,E,p_2,\psi_2,\cdots,p_N,\psi_N\}$. The signature of the manifold is $(\epsilon_t,\epsilon_E,\epsilon_{p_2},\epsilon_{\psi_2},\cdots,\epsilon_{p_N},\epsilon_{\psi_N})$. The metric coefficients will be calculated using Cartan's structural equations as follows. From \eqref{geomN3} the geometric drift follows by integrating the $2$-form
\begin{equation}\label{geom3N}
    d\,\widetilde{\alpha}=d\,\widetilde{\alpha}_t + \sum_{j=2}^{N}d\,\widetilde{\alpha}_j\,,
\end{equation}
where $d\,\widetilde{\alpha}_t=-\frac{1}{\mathsf{A}} dE\wedge\,dt$, and $d\,\widetilde{\alpha}_j=-\frac{1}{\mathsf{A}} dp_j\wedge\, d\psi_j$. The associated $1$-forms are 
\begin{equation}\label{1formsNN}
   \widetilde{\alpha}_t=-\frac{E}{\mathsf{A}} dt\,,\qquad 
   \widetilde{\alpha}_j=-\frac{p_j}{\mathsf{A}} d\psi_j\,,\qquad j=2, \cdots, N\,.
\end{equation}
Drawing on Cartan's structural equations, the $2$-forms $d\,\widetilde{\alpha}_t$ and $d\,\widetilde{\alpha}_j$ are interpreted as the only non-zero curvature $2$-forms of the $2 N$-dimensional shape manifold~$\mathcal{B}$. We relabel the pair $(t,E)$ as $(p_1,\psi_1)$ so that  $d\,\widetilde{\alpha}$ is written as
\begin{equation}\label{geom3N}
    d\,\widetilde{\alpha}=\sum_{j=1}^{N}d\,\widetilde{\alpha}_j\,,
\end{equation}
where we set $\widetilde{\alpha}_1=\widetilde{\alpha}_t$. Comparing with \eqref{geom3} and \eqref{connformalpha}, $\widetilde{\alpha}_j$ and $d\,\widetilde{\alpha}_j$ can be interpreted as the connection and curvature forms of a free $(N+1)$-rotor. Thus, we can use the results we obtained in \S\ref{Sec:metricNpendulum}. For the forced elastic $N$-rotor, the shape manifold~$\mathcal{B}$ has dimension $2 N$. It is reducible since it is the product manifold of $N$ submanifolds (hyper-planes) $\mathcal{B}_j$ with coordinate charts $\{\psi_j,p_j\},\,j=1,\hdots N$. From \eqref{metricN}, the metric of $\mathcal{B}$ is written as
\begin{equation}\label{metricNN}
  \mathbf{G}=\mathbf{G}_1\times\hdots\times\mathbf{G}_N=\sum_{j=1}^N \left[\frac{\mathrm{sgn}(\mathsf{A})}{p_j^2+\mu_j^2}\,dp_j\otimes d p_j+\frac{p_j^2+\mu_j^2}{\mathsf{A}^2}\,d\psi_j\otimes d\psi_j\right]\,, 
\end{equation}
where $\mu_j$ are arbitrary parameters. Since $p_1=t$ and $\psi_1=E$, then  $G_{p_1}=G_t$ and $G_{\psi_1}=G_E$ and the intrinsic metric of each submanifold $\mathcal{B}_j$ follows from \eqref{metric3} as
\begin{equation}\label{metric4a}
	\mathbf{G}_{1}=\frac{\mathrm{sgn}(\mathsf{A})}{E^2+\mu_t^2}\,dE\otimes  dE+\frac{E^2+\mu_t^2}{\mathsf{A}^2}\,dt\otimes dt\,, 
\end{equation}
and
\begin{equation}\label{metric4b}
	\mathbf{G}_{j}=\frac{\mathrm{sgn}(\mathsf{A})}{p_j^2+\mu_j^2}\,dp_j\otimes d p_j
	+\frac{p_j^2+\mu_j^2}{\mathsf{A}^2}\,d\psi_j\otimes d\psi_j\,,\qquad j=2,\cdots, N\,.
\end{equation}
Then 
\begin{equation}\label{metricN4a}
  \mathbf{G}=\frac{\mathrm{sgn}(\mathsf{A})}{E^2+\mu_1^2}\,dE\otimes  dE+\frac{E^2+\mu_1^2}{\mathsf{A}^2}\,dt\otimes dt+\sum_{j=2}^N\,\frac{\mathrm{sgn}(\mathsf{A})}{p_j^2+\mu_j^2}\,dp_j\otimes d p_j+\frac{p_j^2+\mu_j^2}{\mathsf{A}^2}\,d\psi_j\otimes d\psi_j\,. 
\end{equation}
Similar to that of the free elastic $N$-rotor the shape manifold is the product manifold of Robertson-Walker spacetime universes ($\mathsf{A}<0$) or hyperbolic planes  ($\mathsf{A}>0$). 

The geometric drift follows by integrating the $2$-form
\begin{equation}
    d\,\widetilde{\alpha}= \mathcal{R}_{t}^{E}(\mathbf{e}_E,\mathbf{e}_t)+\sum_{j=2}^N\,\mathcal{R}_{\psi_j}^{p_j}(\mathbf{e}_{p_j},\mathbf{e}_{\psi_j})\,,
\end{equation}
which is the sum of the curvature  $2$-forms of each submanifold $\mathcal{B}_j$, that is 
\begin{equation} \label{geom2}
\begin{aligned}
	\theta_{\text{geom}} &=\int_{S(\gamma)} d\,\widetilde{\alpha}=\int_{S(\gamma)}\left[\mathcal{R}_{t}^{E}(\mathbf{e}_E,\mathbf{e}_t)+\sum_{j=2}^N\,\mathcal{R}_{\psi_j}^{p_j}(\mathbf{e}_{p_j},\mathbf{e}_{\psi_j})\right]\\
 &=-\int_{S(\gamma)}\left[\frac{1}{\mathsf{A}} dE\wedge\, dt+\sum_{j=2}^{N}\frac{1}{\mathsf{A}} dp_j\wedge\, d\psi_j\right]\,,
\end{aligned}
\end{equation}
where each term is both the oriented area and curvature of the projected path $\gamma$ on the hyper-planes $\mathcal{B}_j$ and $\mathcal{B}_t$ with coordinates $\{\psi_j,p_j\}$ and $\{t,E\}$, respectively.

\begin{remark}[Metric Uniqueness]
\textcolor{black}{Similarly to the $N$-rotor problem~(see Remark~\ref{Nrotoruniquemetric}), a unique metric  can be defined by matching the symplectic forms~$\beta_t=-\mathsf{A}\,\widetilde{\alpha}_t=E\,dt$ and $\beta_j=-\mathsf{A}\,\widetilde{\alpha}_j=p_j\, d\psi_j$ from \eqref{1formsNN} of the reduced dynamics on~$\mathcal{B}$ with the connection $1$-forms of the base manifold. Such a matching equips $\mathcal{B}$ with the following  pseudo-Riemannian metric
\begin{equation}\label{metricN4a2}
  \mathbf{G}=-\frac{1}{E^2+\mu_1^2}\,dE\otimes  dE+(E^2+\mu_1^2)\,dt\otimes dt+\sum_{j=2}^N\,\left[-\frac{1}{p_j^2+\mu_j^2}\,dp_j\otimes d p_j+(p_j^2+\mu_j^2)\,d\psi_j\otimes d\psi_j\right] \,,
\end{equation}
which is a disguised metric of a multi-universe of $2$D Robertson-Walker spacetimes, $\forall \mu_j\in \mathbb{R}$, $j=1,\hdots,N$. As a result, the geometric phase is directly proportional to curvature, with a constant of proportionality equal to $-\frac{1}{\mathsf{A}}$, see \eqref{geomN3}.}
\end{remark}

\section{Conclusions} \label{Sec:Conclusions}

We investigated the geometric phases of nonlinear elastic $N$-rotors with continuous rotational symmetry in the Hamiltonian framework. The geometric structure of the phase space is a principal fiber bundle, i.e., a base, or shape manifold $\mathcal{B}$, and fibers $\mathcal{F}$ along the symmetry direction attached to it. The connection and curvature forms of the shape manifold are defined by the symplectic structure of the Hamiltonian dynamics. Then, Cartan's moving frames provide the means to derive an intrinsic metric structure for $\mathcal{B}$. This characterizes the kinematically admissible shape deformations of the $N$-rotors. An orbit on $\mathcal{B}$ is a succession of  infinitesimal changes in the shape of the mechanical system from an initial configuration to another. If the mechanical system returns to its initial shape, the orbit is closed and the area~(or curvature) spanned by it measures the induced geometric rotation drift. 
We first studied the geometric phase of a nonlinear elastic double rotor that conserves the total angular momentum $\mathsf{A}$. 
\textcolor{black}{The shape manifold is endowed with two distinct metrics that are compatible with the geometric phase, which depends on the convention used to define the sign of the total angular momentum as either positive or negative, e.g., counterclockwise or clockwise, respectively, or viceversa. If $\mathsf{A}<0$} is chosen, we found that the metric is pseudo-Riemannian and \textcolor{black}{the shape manifold is a $2$D~section of an $4$D~expanding spacetime universe described by the Robertson-Walker metric with positive curvature, and referred to as a $2$D Robertson-Walker spacetime.} \textcolor{black}{If one chooses~$\mathsf{A}>0$}, the shape manifold is the hyperbolic plane $\mathbb{H}^2$ with negative curvature.
\textcolor{black}{A unique metric  can be defined by matching the symplectic form of the reduced dynamics with the curvature form of the shape manifold~$\mathcal{B}$.} 

We next generalized these results to nonlinear elastic $N$-rotors. We found that the associated shape manifold~$\mathcal{B}$ is reducible since it is the product manifold of $N-1$ hyperbolic planes $\mathbb{H}^2$~($\mathsf{A}>0$), or $2$D~Robertson-Walker spacetimes~($\mathsf{A}<0$), \textcolor{black}{ depending on the convention used to define the rotation sign of the total angular momentum.} We then considered elastic $N$-rotors subject to time-dependent self-equilibrated moments. The geometric phase is studied in the extended autonomous Hamiltonian framework. \textcolor{black}{The $(N + 1)$-dimensional shape manifold of the extended
autonomous system has a structure similar to that of the $N$-dimensional shape manifold of free elastic
rotors. Similarly to the double rotor, a unique metric for the $N$-rotors can be defined.} 

\textcolor{black}{The two metrics depend on the sign of $A$ and are both compatible with the geometric phase, which is evaluated by the same $2$-form given by the sum of the sectional curvature forms of $\mathcal{B}$.}
The intrinsic metric allows one to quantify the similarity of a shape $S_1$, or point on $\mathcal{B}$, to another point, or shape $S_2$, by measuring the intrinsic geodesic distance between the two points in terms of curvature, or induced geometric phase. The Euclidean metric would give misleading shorter distances between the two shapes. This is because it is not an intrinsic structure that follows from the dynamics. Thus, low-momentum shapes are far apart from high-momentum shapes. If $\mathsf{A}<0$, the shape manifold is a $2$D~expanding spacetime universe and the two different shapes are red-shifted and are far apart from each other. If $\mathsf{A}>0$, the shape manifold has the character of the hyperbolic plane and the two shapes appear far apart as the difference of their momenta becomes larger. The intrinsic distance between shapes is relevant for measuring how close an orbit is to the stable/unstable submanifolds of fixed points of the dynamics on the shape manifold. 

In future work, we will use Cartan's moving frames to derive an intrinsic metric for the shape manifold of the Navier-Stokes turbulence with continuous translational symmetry, or turbulent channel flows~\citep{Fedele2015}. To unveil the \textit{shape of turbulence} one needs to quotient out the translation symmetry of the Navier-Stokes equations. This can be achieved, for example, by means of a physically meaningful slice or chart representation of the quotient space or shape manifold~\citep{BudanurPRL,Cvitanovic2012,Fedele2015}.  
To measure how close one vortical shape is to another, the standard Euclidean metric is typically used. An important conclusion of our present study is that the similarities of shapes should be measured by a metric intrinsic to the shape manifold. Other non-intrinsic distances are misleading as they do not account for the curvature, or induced geometric phase.

\section*{Acknowledgement}

FF expresses gratitude to Cristel Chandre and Matthew Golden for insightful discussions on Hamiltonian systems and General Relativity. This work was partially supported by NSF -- Grant No. CMMI 1939901, and  ARO Grant No. W911NF-18-1-0003.

\bibliographystyle{plainnat}
\bibliography{ref,ref2,ref3,geometric_phases}

\begin{thebibliography}{38}
\providecommand{\natexlab}[1]{#1}
\providecommand{\url}[1]{\texttt{#1}}
\expandafter\ifx\csname urlstyle\endcsname\relax
  \providecommand{\doi}[1]{doi: #1}\else
  \providecommand{\doi}{doi: \begingroup \urlstyle{rm}\Url}\fi

\bibitem[Aharonov and Anandan(1987)]{Aharonov_Anandan}
Y.~Aharonov and J.~Anandan.
\newblock Phase change during a cyclic quantum evolution.
\newblock \emph{Physical Review Letters}, 58:\penalty0 1593--1596, 1987.

\bibitem[Anandan(1991)]{Anandan_metric}
J.~Anandan.
\newblock A geometric approach to quantum mechanics.
\newblock \emph{Foundations of Physics}, 21:\penalty0 1265--1284, 1991.

\bibitem[Banner et~al.(2014)Banner, Barthelemy, Fedele, Allis, Benetazzo, Dias,
  and Peirson]{Banner_PRL2014}
L.~Banner, M.\, X.~Barthelemy, F.~Fedele, M.~Allis, A.~Benetazzo, F.~Dias, and
  L.~Peirson, W.\.
\newblock Linking reduced breaking crest speeds to unsteady nonlinear water
  wave group behavior.
\newblock \emph{Physical Review Letters}, 112:\penalty0 114502, 2014.

\bibitem[Berry(1984)]{Berry1984}
M.~V. Berry.
\newblock Quantal phase factors accompanying adiabatic changes.
\newblock \emph{Proceedings of the Royal Society of London A}, 392\penalty0
  (1802):\penalty0 45--57, 1984.

\bibitem[Berry(1990)]{Berry1990}
M.~V. Berry.
\newblock Anticipations of the geometric phase.
\newblock \emph{Physics Today}, 43\penalty0 (12):\penalty0 34--40, 1990.

\bibitem[Besse(1987)]{Besse:Besse1987}
A.~L. Besse.
\newblock \emph{Einstein Manifolds}.
\newblock Springer-Verlag, Berlin, Heidelberg, New York, 1987.

\bibitem[Budanur et~al.(2015)Budanur, Cvitanovi\'c, Davidchack, and
  Siminos]{BudanurPRL}
N.~B. Budanur, P.~Cvitanovi\'c, R.~L. Davidchack, and E.~Siminos.
\newblock Reduction of {SO}(2) symmetry for spatially extended dynamical
  systems.
\newblock \emph{Physical Review Letters}, 114:\penalty0 084102, 2015.

\bibitem[Carroll(2003)]{Carroll2003spacetime}
S.~Carroll.
\newblock \emph{Spacetime and Geometry: An Introduction to General Relativity}.
\newblock Benjamin Cummings, 2003.

\bibitem[Cvitanovi\'c et~al.(2012)Cvitanovi\'c, Borrero-Echeverry, Carroll,
  Robbins, and Siminos]{Cvitanovic2012}
P.~P. Cvitanovi\'c, D.~Borrero-Echeverry, K.~M. Carroll, B.~Robbins, and
  E.~Siminos.
\newblock {Cartography of high-dimensional flows: {A} visual guide to sections
  and slices.}
\newblock \emph{Chaos}, 22:\penalty0 047506, Dec. 2012.

\bibitem[Fedele(2014)]{FedeleEPL2014}
F.~Fedele.
\newblock Geometric phases of water waves.
\newblock \emph{Europhysics Letters}, 107\penalty0 (69001), 2014.

\bibitem[Fedele et~al.(2015)Fedele, Abessi, and Roberts]{Fedele2015}
F.~Fedele, O.~Abessi, and P.~J. Roberts.
\newblock Symmetry reduction of turbulent pipe flows.
\newblock \emph{Journal of Fluid Mechanics}, 779:\penalty0 390--410, 9 2015.

\bibitem[Fedele et~al.(2020)Fedele, Banner, and Barthelemy]{fedele2020crest}
F.~Fedele, M.~L. Banner, and X.~Barthelemy.
\newblock Crest speeds of unsteady surface water waves.
\newblock \emph{Journal of Fluid Mechanics}, 899, 2020.

\bibitem[Fedele et~al.(2023)Fedele, Suryanarayana, and Yavari]{Fedele2023}
F.~Fedele, P.~Suryanarayana, and A.~Yavari.
\newblock On the effective dynamic mass of mechanical lattices with
  microstructure.
\newblock \emph{Journal of the Mechanics and Physics of Solids}, 179:\penalty0
  105393, 2023.

\bibitem[Garrison and Chiao(1988)]{Garrison_GeomPhases}
J.~C. Garrison and R.~Y. Chiao.
\newblock Geometrical phases from global gauge invariance of nonlinear
  classical field theories.
\newblock \emph{Physical Review Letters}, 60:\penalty0 165--168, Jan 1988.

\bibitem[Golgoon and Yavari(2018)]{Golgoon2018}
A.~Golgoon and A.~Yavari.
\newblock Line and point defects in nonlinear anisotropic solids.
\newblock \emph{Zeitschrift f{\"u}r angewandte Mathematik und Physik},
  69:\penalty0 1--28, 2018.

\bibitem[Gordeeva et~al.(2010)Gordeeva, Pan’zhenskii, and
  Stepanov]{Gordeeva2010}
I.~Gordeeva, V.~Pan’zhenskii, and S.~Stepanov.
\newblock Riemann--{C}artan manifolds.
\newblock \emph{Journal of Mathematical Sciences}, 169\penalty0 (3):\penalty0
  342--361, 2010.

\bibitem[Hannay(1985)]{Hannay}
J.~H. Hannay.
\newblock Angle variable holonomy in adiabatic excursion of an integrable
  hamiltonian.
\newblock \emph{Journal of Physics A: Mathematical and General}, 18\penalty0
  (2):\penalty0 221, 1985.

\bibitem[Hehl and Obukhov(2003)]{Hehl2003}
F.~W. Hehl and Y.~N. Obukhov.
\newblock \emph{Foundations of Classical Electrodynamics: Charge, Flux, and
  Metric}, volume~33.
\newblock Springer Science \& Business Media, 2003.

\bibitem[Hernández-Garduño and Shashikanth(2018)]{HernandezGarduno2018}
A.~Hernández-Garduño and B.~N. Shashikanth.
\newblock Reconstruction phases in the planar three- and four-vortex problems.
\newblock \emph{Nonlinearity}, 31\penalty0 (3):\penalty0 783--814, 2018.

\bibitem[Joyce(2007)]{Joyce2007}
D.~D. Joyce.
\newblock \emph{Riemannian Holonomy Groups and Calibrated Geometry}, volume~12.
\newblock OUP Oxford, 2007.

\bibitem[Marsden et~al.(1990)Marsden, Montgomery, and Ratiu]{Marsden1990}
J.~E. Marsden, R.~Montgomery, and T.~S. Ratiu.
\newblock \emph{Reduction, Symmetry, and Phases in Mechanics}, volume 436.
\newblock American Mathematical Society, 1990.

\bibitem[Misner et~al.(1973)Misner, Thorne, and Wheeler]{GRAVITATION}
C.~Misner, K.~Thorne, and J.~Wheeler.
\newblock \emph{{Gravitation}}.
\newblock W. Freeman, 1973.

\bibitem[O'Neill(2014)]{ONeill2014}
B.~O'Neill.
\newblock \emph{The geometry of Kerr Black Holes}.
\newblock Courier Corporation, 2014.

\bibitem[Pancharatnam(1956)]{Pancharatnam}
S.~Pancharatnam.
\newblock Generalized theory of interference, and its applications.
\newblock \emph{Proceedings of the Indian Academy of Sciences - Section A},
  44\penalty0 (5):\penalty0 247--262, 1956.

\bibitem[Provost and Vallee(1980)]{ProvostVallee}
J.~P. Provost and G.~Vallee.
\newblock Riemannian structure on manifolds of quantum states.
\newblock \emph{Communications in Mathematical Physics}, 76:\penalty0 289--301,
  1980.

\bibitem[Samuel and Bhandari(1988)]{Bhandari}
J.~Samuel and R.~Bhandari.
\newblock General setting for {B}erry's phase.
\newblock \emph{Physical Review Letters}, 60:\penalty0 2339--2342, 1988.

\bibitem[Shapere and Wilczek(1987)]{Shapere1987}
A.~Shapere and F.~Wilczek.
\newblock Self-propulsion at low reynolds number.
\newblock \emph{Physical Review Letters}, 58\penalty0 (20):\penalty0 2051,
  1987.

\bibitem[Shapere and Wilczek(1989)]{Shapere1}
A.~Shapere and F.~Wilczek.
\newblock Geometry of self-propulsion at low {R}eynolds number.
\newblock \emph{Journal of Fluid Mechanics}, 198:\penalty0 557--585, 1 1989.

\bibitem[Shashikanth and Marsden(2003)]{leapfrogvortexShashikanthMarsden2003}
B.~N. Shashikanth and J.~E. Marsden.
\newblock Leapfrogging vortex rings: Hamiltonian structure, geometric phases
  and discrete reduction.
\newblock \emph{Fluid Dynamics Research}, 33\penalty0 (4):\penalty0 333--356,
  oct 2003.

\bibitem[Simon(1983)]{Simon1983}
B.~Simon.
\newblock Holonomy, the quantum adiabatic theorem, and {B}erry's phase.
\newblock \emph{Physical Review Letters}, 51\penalty0 (24):\penalty0 2167,
  1983.

\bibitem[Sternberg(1999)]{Sternberg1999}
S.~Sternberg.
\newblock \emph{Lectures on Differential Geometry}, volume 316.
\newblock American Mathematical Society, 1999.

\bibitem[Sternberg(2013)]{Sternberg2013}
S.~Sternberg.
\newblock \emph{Curvature in Mathematics and Physics}.
\newblock Courier Corporation, 2013.

\bibitem[Wilczek and Shapere(1989)]{Wilczek_book}
F.~Wilczek and A.~Shapere.
\newblock \emph{Geometric Phases in Physics}.
\newblock World Scientific, 1989.
\newblock \doi{10.1142/0613}.

\bibitem[Yavari(2016)]{Yavari2016Dispiration}
A.~Yavari.
\newblock On the wedge dispiration in an inhomogeneous isotropic nonlinear
  elastic solid.
\newblock \emph{Mechanics Research Communications}, 78:\penalty0 55--59, 2016.

\bibitem[Yavari and Goriely(2012{\natexlab{a}})]{Yavari2012a}
A.~Yavari and A.~Goriely.
\newblock Riemann--{C}artan geometry of nonlinear dislocation mechanics.
\newblock \emph{Archive for Rational Mechanics and Analysis}, 205:\penalty0
  59--118, 2012{\natexlab{a}}.

\bibitem[Yavari and Goriely(2012{\natexlab{b}})]{Yavari2012b}
A.~Yavari and A.~Goriely.
\newblock Weyl geometry and the nonlinear mechanics of distributed point
  defects.
\newblock \emph{Proceedings of the Royal Society A: Mathematical, Physical and
  Engineering Sciences}, 468\penalty0 (2148):\penalty0 3902--3922,
  2012{\natexlab{b}}.

\bibitem[Yavari and Goriely(2013)]{Yavari2013}
A.~Yavari and A.~Goriely.
\newblock Riemann--{C}artan geometry of nonlinear disclination mechanics.
\newblock \emph{Mathematics and Mechanics of Solids}, 18\penalty0 (1):\penalty0
  91--102, 2013.

\bibitem[Yavari and Goriely(2014)]{Yavari2014}
A.~Yavari and A.~Goriely.
\newblock The geometry of discombinations and its applications to semi-inverse
  problems in anelasticity.
\newblock \emph{Proceedings of the Royal Society A: Mathematical, Physical and
  Engineering Sciences}, 470\penalty0 (2169):\penalty0 20140403, 2014.

\end{thebibliography}

\end{document}